\documentclass[11pt]{article}

\pdfoutput=1
\usepackage{jheppub}

\makeatletter
\def\@fpheader{$\,$}
\makeatother

\usepackage{amsmath}
\usepackage{amsfonts}
\usepackage{wasysym}
\usepackage{graphicx}
\usepackage{comment}
\usepackage{tensor}
\usepackage{cancel}
\usepackage[svgnames]{xcolor}

\def\filetype{pdf}
\def\path{}

\allowdisplaybreaks[1]

\title{Entanglement entropy of evolving wormholes}
\author[a]{Ben Kain}
\author[b]{and Hong Y.~Ling}
\affiliation[a]{Department of Physics, College of the Holy Cross, Worcester, Massachusetts 01610, USA}
\affiliation[b]{Department of Physics and Astronomy, Rowan University, Glassboro, New Jersey 08028, USA}

\abstract{
We compute the entanglement entropy of a discretized scalar field in the fixed spacetime background of an evolving wormhole. Evolving wormholes are a class of time dependent and spherically symmetric wormhole geometries. In both expanding and contracting models, we find dynamically generated entanglement. Asymptotically far from the wormhole throat, the entanglement entropy is well described by the leading term in the area law, although its coefficient can depend sensitively on the value of the UV cutoff. As the UV cutoff is decreased, the coefficient becomes progressively less sensitive to the cutoff, providing numerical evidence that it approaches a cutoff-independent value in the continuum limit. Near the throat, we find a double-dip structure in the entanglement entropy which cannot be described by the leading and logarithmic contributions to the area law.}

\begin{document}

\maketitle
\flushbottom


\section{Introduction}

\label{sec:intro}

In their seminal work, Morris and Thorne studied time independent and spherically symmetric wormholes \cite{Morris:1988cz}. They showed that for certain choices of the functions used to parameterize the static metric, wormholes could be traversable, but that any matter which could be a source for the wormhole geometry would violate the null energy condition.

Evolving wormholes are a class of wormhole geometries which generalize the Morris and Thorne metric by including a time dependence \cite{Roman:1992xj, Kar:1994tz, Kar:1995ss, Anchordoqui:1997du}. A significant property of evolving wormholes is that for some choices of the metric functions, the wormhole can satisfy the weak energy condition, and hence can also satisfy the null energy condition. Evolving wormholes, unlike their static counterparts studied by Morris and Thorne, do not necessarily require exotic matter. 

It has been known for some time that the relativistic vacuum exhibits spatial entanglement \cite{Bombelli:1986rw, Srednicki:1993im}. In a spherically symmetric system, a lattice can be introduced by discretizing the radial axis. The vacuum state for a scalar field propagating on the lattice can be described with a density matrix. By tracing out degrees of freedom on the lattice, which is equivalent to tracing out some portion of space, the entanglement entropy can be computed as a measure of the spatial entanglement of the vacuum. This was first done in flat space \cite{Bombelli:1986rw, Srednicki:1993im}, where it was shown that the entanglement entropy scales with the area of the boundary, where the spherical boundary lies between the region traced out and the region that is not. Subsequent studies showed that this continues to be the case for static curved spacetimes \cite{Mukohyama:1997aq, Das:2007mj, Das:2008sy}.

While substantial work has been done studying entanglement entropy between two spatial regions in static spacetimes, substantially less has been done for time dependent spacetimes \cite{Boutivas:2023ksg, Katsinis:2023hqn, Boutivas:2023mfg, Belfiglio:2025hzo}. In fact, there appears to be evidence that in a time dependent spacetime, the entanglement entropy may not scale with the area of the boundary \cite{Belfiglio:2025hzo}. Computing the entanglement entropy of a time dependent spacetime is nontrivial. Recent developments have generalized Srednicki's method for computing the entanglement entropy of a discretized scalar field \cite{Srednicki:1993im} to time dependent spacetimes for which the metric takes the FLRW form \cite{Boutivas:2023ksg, Katsinis:2023hqn}. This is precisely the form used for evolving wormholes.

In the standard non-wormhole calculation \cite{Srednicki:1993im}, one introduces a UV cutoff, which acts as the spacing between lattice points. At the origin of the spherical coordinate system, the areal radius is zero, which is necessarily smaller than the UV cutoff. As a consequence, the computation of the entanglement entropy should not be trusted too close to the origin. One pleasant feature of computing the entanglement entropy of a wormhole, is that the areal radius has a minimum nonzero value, marking the wormhole throat. As long as one considers wormhole models in which the wormhole throat radius is much larger than the UV cutoff, the entanglement entropy calculation can be trusted throughout the spacetime.

In this work, we compute the entanglement entropy of a discretized scalar field in the fixed spacetime background of an evolving wormhole. One goal of this work is to determine how the entanglement entropy changes as the wormhole geometry evolves. We find, in both expanding and contracting models, that entanglement entropy is dynamically generated. A second goal is to determine if the entanglement entropy obeys the area law. The initial static configuration exhibits the expected area law behavior. As the evolution proceeds, we find that asymptotically far from the wormhole throat the entanglement entropy scales with the leading contribution to the area law, but that near the wormhole throat, both the leading and logarithmic contributions are insufficient for describing the entanglement entropy, raising questions about the subleading behavior of the entanglement entropy in this region.

In section \ref{sec:evolving wormholes}, we briefly review evolving wormholes, focusing on those properties relevant for computing the entanglement entropy. In section \ref{sec:EE}, we give a careful derivation of the equations for computing the entanglement entropy,  with the step-by-step procedure for the computation given in section \ref{sec:procedure}. Since section \ref{sec:EE} is lengthy, and also to help the reader less interested in the details, we have written section \ref{sec:procedure} in such a way that it can be read without having read the previous subsections. In section \ref{sec:EE evolving wormholes}, we compute the entanglement entropy for two specific evolving wormhole models and analyze the results. We conclude in section \ref{sec:conclusion}. Throughout, we set $c = G = \hbar = 1$.


\section{Evolving wormholes}
\label{sec:evolving wormholes}

Morris and Thorne studied time independent and spherically symmetric wormholes described by the metric \cite{Morris:1988cz}
\begin{equation} \label{MT metric}
ds^2 = - e^{2\Phi(r)} dt^2 + \frac{dr^2}{1-h(r)/r} + r^2 d\Omega^2,
\end{equation}
where $\Phi(r)$ and $h(r)$, known respectively as the redshift function and the shape function, parametrize the metric and $d\Omega^2 = d\theta^2 + \sin^2\theta \, d\phi^2$. Morris and Thorne showed that for certain choices of $\Phi(r)$ and $h(r)$, the metric in (\ref{MT metric}) describes a traversable static wormhole. They further showed that any matter which could be a source for the wormhole geometry would violate the null energy condition.

Evolving wormholes are a class of wormhole geometries which generalize the Morris and Thorne metric by including a time dependence \cite{Roman:1992xj, Kar:1994tz, Kar:1995ss, Anchordoqui:1997du}. The evolving wormhole metric takes the form
\begin{equation} \label{evolving metric}
ds^2 = - A^2(t) e^{2\Phi(r)} dt^2 + B^2(t) \left[ \frac{dr^2}{1-h(r)/r} + r^2 d\Omega^2 \right],
\end{equation}
where $A(t)$ and $B(t)$ are time dependent functions and where $A(t)$ is constrained to take one of the following two forms:
\begin{equation} \label{A choices}
A(t) = 1
\quad \text{or} \quad
A(t) = B(t).
\end{equation}
When $A(t) = 1$, $t$ is known as real time and $B(t)$ is a scale factor familiar from studies of cosmology. When $A(t) = B(t)$, $t$ is known as conformal time and $B(t)$ is a conformal factor. The two choices for $A(t)$ in (\ref{A choices}) are related by a simple transformation of the time coordinate,
\begin{equation}
dt = B(\tau) d\tau,
\end{equation}
where, for the purposes of writing down the transformation, $t$ is real time and $\tau$ is conformal time.

The areal radius for the evolving wormhole metric in (\ref{evolving metric}) is $R(t,r) = B(t) r$, where $r$ also serves as the radial coordinate. Since $r$ must be non-negative, this is a poor choice for the radial coordinate since it cannot describe both sides of the wormhole. We introduce a useful new radial coordinate, $x$, defined by
\begin{equation} \label{dx dr}
dx = \pm \frac{dr}{\sqrt{1-h(r)/r}},
\end{equation}
so that the metric becomes
\begin{equation} \label{evolving metric x}
ds^2 = - A^2(t) e^{2\Phi(x)} dt^2 + B^2(t) \left[ dx^2 + r^2(x) d\Omega^2 \right].
\end{equation}
Integrating (\ref{dx dr}), 
\begin{equation} \label{x(r) int}
x(r) = \pm \int_{r_\text{th}} \frac{dr}{\sqrt{1-h(r)/r}}.
\end{equation}
At the wormhole throat, $r$ will have its minimum value, $r_\text{th}$. The integration limit indicates that our convention is to set $x(r_\text{th}) = 0$. With this convention, the upper signs in (\ref{dx dr}) and (\ref{x(r) int}) describe one side of the wormhole and the lower signs describe the other (we will only consider evolving wormholes that are symmetric about the wormhole throat). Our new radial coordinate takes values in the range $-\infty < x < \infty$. We will use the form of the metric in (\ref{evolving metric x}) when we compute the entanglement entropy. Note that the proper distance from the center of the wormhole at $x = 0$ is given by
\begin{equation} \label{proper dist}
\ell(t, x) = B(t)x = \pm B(t)  \int_{r_\text{th}} \frac{dr}{\sqrt{1-h(r)/r}}.
\end{equation}

To show that the metrics in (\ref{evolving metric}) and (\ref{evolving metric x}) describe a wormhole structure, one may construct a 2D embedding surface in flat 3D space which has the same geometry as a constant time and constant angle slice of the metric, as was done in \cite{Roman:1992xj, Kar:1994tz, Kar:1995ss, Anchordoqui:1997du}. Here, we take a different route which makes use of our new radial coordinate. In terms of the new radial coordinate, the areal radius is given by
\begin{equation}
R(t,x) = B(t) r(x).
\end{equation}
A wormhole structure exists for a given time $t$ if $R$ has a positive minimum. We may take $x$-derivatives of $R$ with the help of (\ref{dx dr}). Setting the first $x$-derivative equal to zero leads to
\begin{equation}
h(r) = r.
\end{equation}
An extremum exists at the solution to this equation. Let the solution be $r = r_\text{th}$. For this extremum to be a minimum, the second derivative evaluated at $r = r_\text{th}$ must be positive, which leads to
\begin{equation}
0 < - \frac{dh(r)}{dr} + \frac{h(r)}{r} \Biggr|_{r=r_\text{th}} = - \frac{dh(r)}{dr} \Biggr|_{r=r_\text{th}} + 1
\end{equation}
or 
\begin{equation}
\frac{dh(r)}{dr} \Biggr|_{r=r_\text{th}} < 1.
\end{equation}
We find that whether or not a wormhole structure exists depends only on the choice of the shape function, $h(r)$. Further, if it does exist, it will exist at all times. A remaining constraint on $h(r)$ is that $h(r) \leq r$ for $r \geq r_\text{th}$, so that the proper distance in (\ref{proper dist}) is well-defined. In practice, one may choose a shape function consistent with the $h(r) \leq r$ constraint and then solve $h(r_\text{th}) = r_\text{th}$ for $r_\text{th}$. If $dh/dr$ evaluated at $r = r_\text{th}$ is less than 1, the metric describes a wormhole.

A commonly used and particularly simple choice for $h(r)$ is for it to be constant \cite{Roman:1992xj, Kar:1994tz, Kar:1995ss, Anchordoqui:1997du},
\begin{equation}
h(r) = \text{r}_{\text{th}}.
\end{equation}
This clearly satisfies the conditions for the presence of a wormhole. When we present results for the entanglement entropy, we will do so using this choice for $h(r)$. A constant $h(r)$ allows the integral in (\ref{x(r) int}) to be evaluated analytically, giving
\begin{equation} \label{x(r)}
x = \pm r \left( \sqrt{1 - \frac{r_\text{th}}{r}} 
+ \frac{r_\text{th}}{r} \tanh^{-1} \sqrt{1 - \frac{r_\text{th}}{r}}
\right).
\end{equation}
A commonly used choice for the redshift function is $\Phi = 0$ \cite{Roman:1992xj, Kar:1994tz, Kar:1995ss}. In the next section, we will see that our method for computing the entanglement entropy only works for $\Phi = 0$ and, as such, our results for the entanglement entropy will also use $\Phi = 0$.

As mentioned in the introduction, for some choices of the metric functions, evolving wormholes can satisfy the weak energy condition. The details of this fact do not affect the computation of the entanglement entropy and we will not review them here (for more information, see \cite{Kar:1994tz, Kar:1995ss, KordZangeneh:2025qel}).


\section{Entanglement entropy}
\label{sec:EE}

In this section, we derive the equations and present the procedure for computing the entanglement entropy of a discretized scalar field propagating in a fixed spacetime background with metric
\begin{equation} \label{EE metric}
ds^2 = -A^2(t) dt^2 + B^2(t) \left[b^2(x) dx^2 + r^2(x) d\Omega^2 \right].
\end{equation}
This is the most general time dependent and spherically symmetric metric for which we have been able to compute the entanglement entropy.

We note that the computation of the entanglement entropy does not require the time dependent function $A(t)$ to obey the evolving wormhole constraint in (\ref{A choices}). That said, in the next section when we apply our results to evolving wormholes, we will make use of the constraint. Further, comparing this metric to the evolving wormhole metric in (\ref{evolving metric x}), we find that we can only compute the entanglement entropy for evolving wormholes with $\Phi = 0$.

The derivation of the equations for computing the entanglement entropy is lengthy. The reader uninterested in the details may skip to section \ref{sec:procedure}, where we describe the step-by-step procedure with minimal justification.


\subsection{Discretized Lagrangian}

The Lagrangian for a real massive scalar field is
\begin{equation} \label{phi L}
\begin{split}
L(t) 
&= \int \sqrt{-\det(g_{\mu\nu})} \, 
\left[ -\frac{1}{2} (\nabla_\mu \phi)(\nabla^\mu \phi) - \frac{1}{2} \mu^2 \phi^2 \right]
dx d\theta d\phi
\\
&= \int A B^3 b r^2 
\left[ -\frac{1}{2} (\nabla_\mu \phi)(\nabla^\mu \phi) - \frac{1}{2} \mu^2 \phi^2 \right]
dx d\Omega,
\end{split}
\end{equation}
where $\phi$ is the scalar field, $\mu$ is the mass, and $d\Omega = \sin\theta d\theta d\phi$. As a real field in a spherically symmetric spacetime, $\phi$ may be decomposed in terms of the real spherical harmonics,
\begin{equation} \label{phi Y decomp}
\phi(t,x,\theta,\phi)
= \sum_{\ell = 0}^\infty \sum_{m=-\ell}^\ell
\phi_{\ell m}(t,x) Y_{\ell m}(\theta,\phi),
\end{equation}
where the $Y_{\ell m}$ satisfy the orthonormality relation
\begin{equation} \label{Y ortho}
\int d\Omega \, Y_{\ell m}(\theta,\phi) Y_{\ell' m'}(\theta,\phi) = \delta_{\ell \ell'} \delta_{m m'}.
\end{equation}
Using (\ref{phi Y decomp}) and (\ref{Y ortho}), we can evaluate the integrals over the angular coordinates in (\ref{phi L}), obtaining
\begin{equation} \label{L}
L(t) =  \sum_{\ell = 0}^\infty \sum_{m=-\ell}^\ell
\int dx \, \mathcal{L}_{\ell m}(t,x),
\end{equation}
where the Lagrangian density is
\begin{equation} \label{L lm}
\begin{split}
\mathcal{L}_{\ell m}(t,x)
&= \frac{1}{2} 
\frac{B^3}{A}
b r^2
(\partial_t \phi_{\ell m})^2
-  \frac{1}{2}
 AB \frac{r^2}{b} 
(\partial_x \phi_{\ell m})^2
- \frac{1}{2}
 AB b r^2 
\left[ B^2 \mu^2   
+ \frac{\ell(\ell + 1)}{r^2} \right]  \phi_{\ell m}^2.
\end{split}
\end{equation}
The integrals leading to (\ref{L}) and (\ref{L lm}) are straightforward to evaluate, except for the one containing $\theta$- and $\phi$-derivatives. To evaluate this integral, we performed an integration by parts and used the eigenvalue equation for the spherical harmonics,
\begin{equation}
\frac{1}{\sin\theta}\partial_\theta (\sin\theta \, \partial_\theta Y_{\ell m})
+ \frac{1}{\sin^2\theta} \partial_\phi^2 Y_{\ell m}
= -\ell(\ell+1) Y_{\ell m}.
\end{equation}

The method we are presenting will be able to compute the entanglement entropy if the coefficient of the kinetic term in (\ref{L lm}) is independent of $x$. We therefore move to a new field, $\varphi(t,x)$, defined with normalization
\begin{equation}
\varphi_{\ell m} (t,x) \equiv r(x) \sqrt{b(x)} \, \phi_{\ell m}(t,x).
\end{equation}
In terms of $\varphi(t,x)$, the Lagrangian density is 
\begin{equation} \label{L density}
\begin{split}
\mathcal{L}_{\ell m}(t,x)
&= \frac{1}{2} 
\frac{B^3}{A}
(\partial_t \varphi_{\ell m})^2
- \frac{1}{2} 
 AB \frac{r^2}{b} 
\left[ \partial_x \left( \frac{1}{r\sqrt{b}} \, \varphi_{\ell m} \right) \right]^2
- \frac{1}{2}  
A B
\left[ B^2 \mu^2   
+ \frac{\ell(\ell + 1)}{r^2} \right]  
\varphi_{\ell m}^2.
\end{split}
\end{equation}

The next step is to discretize the radial axis, $x$, which will lead to a discretized scalar field. This introduces a grid or lattice, where $\epsilon$ is the spacing between lattice sites. Physically, $\epsilon$ is interpreted as an ultraviolet cutoff. The points on our lattice are given by
\begin{equation} \label{grid def}
x_j = \left[ j - \frac{1}{2}(N + 1)\right]\epsilon,
\qquad
j = 1,2,\ldots,N,
\end{equation}
where $N$ is the number of lattice sites and is even. It is convenient to write out some of the $x_j$-values:
\begin{equation} \label{discrete x values}
\left(-\frac{N}{2} + \frac{1}{2} \right) \epsilon,
\quad
\left( - \frac{N}{2} + \frac{3}{2} \right) \epsilon,
\quad
\ldots,
\quad
-\frac{1}{2}\epsilon, \, \frac{1}{2}\epsilon,  \,\frac{3}{2}\epsilon,
\quad
\ldots,
\quad
\left( \frac{N}{2} - \frac{1}{2} \right) \epsilon.
\end{equation}
The $x_j$-values are symmetric about $x = 0$ and do not include $x = 0$. The two lattice sites closest to $x = 0$, which correspond to $x = -\epsilon/2$ and $x = +\epsilon/2$ have, respectively, indices
\begin{equation}
j = \frac{N}{2}
\quad \text{and} \quad
j = \frac{N}{2} + 1.
\end{equation}
The spatially-dependent fields are now discretized as
\begin{equation}
\varphi_{\ell m j}(t) \equiv \varphi_{\ell m}(t,x_j),
\qquad
b_j \equiv b(x_j),
\qquad
r_j \equiv r(x_j).
\end{equation}

To discretize the Lagrangian density in (\ref{L density}), we must handle the $x$-derivative. We use the formulas 
\begin{equation} \label{finite diff}
[\partial_x f(x)]_j = \frac{f_{j+1} - f_{j}}{\epsilon},
\qquad
\left[ g(x) \partial_x f(x) \right]_j
= g_{j+1/2} \frac{f_{j+1} - f_{j}}{\epsilon},
\end{equation}
where $f$ and $g$ are arbitrary functions. These formulas are the standard ones used in the literature for $x \geq 0$ \cite{Srednicki:1993im, Mukohyama:1997aq, Das:2007mj, Das:2008sy, Belfiglio:2025hzo}. We have simply extended them to $x < 0$. We note that they are not symmetric about $x = 0$. As a consequence, the values for the entanglement entropy that we compute will not be exactly symmetric about $x = 0$, although the spacetime is symmetric. It is possible to use formulas that are symmetric, but they generally lead to a more complicated construction.

Using the formulas in (\ref{finite diff}), the Lagrangian density can be written 
\begin{equation} \label{discretized L density 1st order}
\begin{split}
2\epsilon^2 \mathcal{L}_{\ell m j}(t)
&= 
\epsilon^2
\frac{B^3(t)}{A(t)}
(\partial_t \varphi_{\ell m j})^2
- A(t) B^3(t) \tilde{\mu}^2   \varphi_{\ell m j}^2
\\
&\qquad
+ A(t)B(t) \left(
\alpha_j \varphi_{\ell m, j+1}^2
+ \beta_j \varphi_{\ell m j}^2
+ 2 \gamma_j \varphi_{\ell m j} \varphi_{\ell m, j+1}  \right),
\end{split}
\end{equation}
where
\begin{equation} \label{alpha beta gamma def}
\begin{split}
\alpha_j &= - \frac{\tilde{r}^2_{j+1/2}}{\tilde{r}^2_{j+1}}
\frac{1}{b_{j+1/2} b_{j+1}} 
\\
\beta_j &=
-\frac{\tilde{r}^2_{j+1/2}}{\tilde{r}_j^2}
\frac{ 1}{b_{j+1/2}b_j} 
-  \frac{\ell(\ell + 1)}{ \tilde{r}_j^2} 
\\
\gamma_j &=
\frac{\tilde{r}^2_{j+1/2}}{\tilde{r}_{j+1} \tilde{r}_j}
\frac{1}{b_{j+1/2}\sqrt{b_{j+1} b_{j}}}
\end{split}
\end{equation}
and
\begin{equation} \label{r mu scaling}
\tilde{r}_j \equiv \frac{r_j}{\epsilon},
\qquad
\tilde{\mu} \equiv \mu \epsilon.
\end{equation}
The Lagrangian density is integrated over the $x$-axis to give the Lagrangian, as can be seen from (\ref{L}). Having discretized the $x$-axis, the integral becomes a sum and
\begin{equation} \label{L sum x}
L(t) =
\sum_{\ell = 0}^\infty \sum_{m=-\ell}^\ell
\epsilon
\sum_{j = 1}^{N} \mathcal{L}_{\ell m j}(t).
\end{equation}
Inserting (\ref{discretized L density 1st order}) into (\ref{L sum x}) gives for the Lagrangian
\begin{equation} \label{Lagrangian discrete 0}
\begin{split}
L(t) 
&= \frac{1}{2 \epsilon}
\sum_{\ell = 0}^\infty \sum_{m=-\ell}^\ell
\left\{
\epsilon^2 
\frac{B^3}{A}
\sum_{j = 1}^{N}
(\partial_t \varphi_{\ell m j})^2
- A B^3 \tilde{\mu}^2  \sum_{j = 1}^{N} \varphi_{\ell m j}^2
\right.
\\
&\qquad
\left.
+ AB \left[ 
\beta_1 \varphi_{\ell m, 1}^2
+ \sum_{j = 2}^{N} \left( \beta_j + \alpha_{j-1} \right)
\varphi_{\ell m j}^2
+ 2  \sum_{j = 1}^{N-1} \gamma_j \varphi_{\ell m j} \varphi_{\ell m, j+1}
\right]
\right\}.
\end{split}
\end{equation}
In writing this form down, we have introduced the standard infrared cutoff,
\begin{equation} \label{BCs}
\varphi_{\ell m, N+1} = 0.
\end{equation}
Also part of the infrared cutoff is that the sum over $j$ in (\ref{L sum x}) does not include $j < 1$. Numerically, we are introducing boundary conditions for the lattice. 

To proceed, we write the Lagrangian in (\ref{Lagrangian discrete 0}) as 
\begin{equation} \label{Lagrangian discrete}
\begin{split}
L(t)= \frac{1}{2\epsilon}
\sum_{\ell = 0}^\infty \sum_{m=-\ell}^\ell
&\Biggl[ \epsilon^2 M(t) \sum_{j =1}^{N} 
(\partial_t \varphi_{\ell m j})^2 
- A(t) B^3(t) \tilde{\mu}^2 
\sum_{j=1}^N \varphi_{\ell m j}^2
\\
&\quad
- A(t)B(t)
\sum_{j = 1}^{N}
\sum_{k = 1}^{N}
\varphi_{\ell m,j} K_{\ell m, jk} \varphi_{\ell m, k}
\Biggr],
\end{split}
\end{equation}
where
\begin{equation} \label{M def}
M(t) = \frac{B^3(t)}{A(t)}.
\end{equation}
The components of $K_{\ell m, jk}$ follow directly from (\ref{Lagrangian discrete 0}). We write them out explicitly below in (\ref{K diag}) and (\ref{K off-diag}). We have succeeding in discretizing the scalar field, arriving at the discrete set of fields $\varphi_{\ell m j}$ which are described by the Lagrangian in (\ref{Lagrangian discrete}).


\subsection{Hamiltonian and quantization}

To canonically quantize the discrete set of scalar fields, we first construct the Hamiltonian. From the Lagrangian in (\ref{Lagrangian discrete}), the canonically conjugate momentum are
\begin{equation}
\pi_{\ell m j}(t)
= \frac{\partial L(t)}{\partial (\partial_t \varphi_{\ell m j})}
= \epsilon  M(t) \partial_t \varphi_{\ell m j},
\end{equation}
and the Hamiltonian is
\begin{equation} \label{H}
H(t) = \sum_{\ell = 0}^\infty \sum_{m=-\ell}^\ell H_{\ell m}(t),
\end{equation}
where 
\begin{equation} \label{H density}
\begin{split}
H_{\ell m}(t) = \frac{1}{\epsilon}
\sum_{j=1}^N
&\Biggl[ \frac{\pi_{\ell m j}^2 }{2M(t)}
+ \frac{1}{2} A(t) B^3(t) \tilde{\mu}^2
\varphi_{\ell m j}^2
+ \frac{1}{2} A(t) B(t)
\sum_{k = 1}^{N}
\varphi_{\ell m,j} K_{\ell m, jk} \varphi_{\ell m, k}
\Biggr].
\end{split}
\end{equation}

We quantize by elevating the fields $\varphi_{\ell m j}$ and $\pi_{\ell m j}$ to operators which satisfy the canonical commutation relation
\begin{equation}
[\varphi_{\ell m j}, \pi_{\ell' m' j'}] = i \delta_{\ell \ell'} \delta_{m m'} \delta_{j j'}.
\end{equation}
The Hamiltonian operator is then used in the time dependent functional Schr\"odinger equation \cite{Das:2008sy, Hatfield:2019sox},
\begin{equation} \label{Schrodinger 1}
i \frac{d\Psi[\varphi]}{dt} = H \Psi[\varphi],
\end{equation}
or
\begin{equation} \label{Schrodinger 2}
i  \frac{d\Psi[\varphi]}{d\tilde{t}}
= 
\sum_{\ell = 0}^\infty \sum_{m=-\ell}^\ell
\sum_{j =1}^{N} 
\left[ \frac{\pi_{\ell m j}^2 }{2M(\tilde{t})}
+ \frac{1}{2} A(\tilde{t}) B^3(\tilde{t}) \tilde{\mu}^2
\varphi_{\ell m j}^2
+ \frac{1}{2} A(\tilde{t})B(\tilde{t})
\sum_{k = 1}^{N}
\varphi_{\ell m,j} K_{\ell m, jk} \varphi_{\ell m, k}
\right]
\Psi[\varphi],
\end{equation}
where
\begin{equation} \label{t scaling}
\tilde{t} \equiv \frac{t}{\epsilon}.
\end{equation}
Note that the functional Schr\"odinger equation in (\ref{Schrodinger 2}) is independent of $m$ and different values of $\ell$ do not couple. This suggests that we look for solutions to the related Schr\"odinger equation
\begin{equation} \label{Schrodinger 3}
i  \frac{d\psi_{\ell m}[\varphi]}{d\tilde{t}}
= 
\sum_{j =1}^{N} 
\left[ \frac{\pi_{\ell m j}^2 }{2M(\tilde{t})}
+ \frac{1}{2} A(\tilde{t}) B^3(\tilde{t}) \tilde{\mu}^2
\varphi_{\ell m j}^2
+ \frac{1}{2} A(\tilde{t})B(\tilde{t})
\sum_{k = 1}^{N}
\varphi_{\ell m,j} K_{\ell m, jk} \varphi_{\ell m, k}
\right]
\psi_{\ell m}[\varphi],
\end{equation}
from which we can form the solution we are after as
\begin{equation} \label{Psi psi lm}
\Psi[\varphi] = \prod_{\ell = 0}^\infty \prod_{m=-\ell}^\ell \psi_{\ell m}[\varphi].
\end{equation}

Before proceeding, it is useful to simplify the notation. This will make the problem we are trying to solve clearer. We drop the subscripted $\ell m$ and replace the scalar and momentum fields with $z_j$ and $p_j$. With this change in notation, we aim to solve the time dependent Schr\"odinger equation
\begin{equation}
i \frac{d\psi}{d\tilde{t}} = H \psi,
\end{equation}
where the Hamiltonian is
\begin{equation} \label{H simple}
H
= 
\sum_{j =1}^{N} 
\left[ \frac{p_{j}^2 }{2M(\tilde{t})}
+ \frac{1}{2} A(\tilde{t}) B^3(\tilde{t}) \tilde{\mu}^2
z_j^2
+ \frac{1}{2} A(\tilde{t})B(\tilde{t})
\sum_{k = 1}^{N}
z_j K_{jk} z_k
\right].
\end{equation}
It is now clear that we are solving the Sch\"odinger equation for $N$ coupled harmonic oscillators with time dependent frequencies and a common time dependent mass.

The matrix $K_{jk}$ is real, symmetric, time independent, and positive definite. It can be diagonalized by a time independent orthogonal matrix, $U$. Our convention is
\begin{equation}
K = U^T K_D U,
\end{equation}
where $K_D$ is the diagonal matrix. Since $K$ is positive definite, we can write
\begin{equation}
(K_D)_{jk} = \Gamma_j^2 \delta_{jk},
\end{equation}
where $\Gamma_j^2$ are the eigenvalues of $K$. We now adopt a field-space representation, where $p_j = -id/dz_j$, and define
\begin{equation} \label{y U z}
\mathbf{y}= U \mathbf{z}.
\end{equation}
The Hamiltonian is now
\begin{equation} \label{uncoupled H}
H = \sum_{j=1}^N \left[ -\frac{1}{2 M(\tilde{t})}\frac{d^2}{dy_j^2}
+ \frac{1}{2} M(\tilde{t}) \omega_j^2(\tilde{t}) y_j^2 \right],
\end{equation}
where
\begin{equation} \label{M w eqs}
M(\tilde{t}) = \frac{B^3(\tilde{t})}{A(\tilde{t})},
\qquad
\omega_i^2(\tilde{t}) = \frac{A^2(\tilde{t})}{B^2(\tilde{t})} \left[ \Gamma_i^2 + B^2(\tilde{t}) \tilde{\mu}^2 \right].
\end{equation}
We have succeeded in diagonalizing the Hamiltonian, which now describes $N$ uncoupled harmonic oscillators $y_j$ with time dependent frequencies $\omega_j(\tilde{t})$ and a common time dependent mass $M(\tilde{t})$. The final form of the Schr\"odinger equation is then
$i d\psi_j(\tilde{t},y_j)/d\tilde{t} = H_j \psi_j(\tilde{t},y_j)$,
where 
\begin{equation} \label{H j}
H_j = -\frac{1}{2 M(\tilde{t})}\frac{d^2}{dy_j^2}
+ \frac{1}{2} M(\tilde{t}) \omega_j^2(\tilde{t}) y_j^2
\end{equation}
and 
\begin{equation} \label{psi psi j}
\psi = \prod_{j=1}^N \psi_j.
\end{equation}


\subsection{Density matrix}

Our goal is to compute the entanglement entropy for the vacuum. In the functional Schr\"odinger formalism, the vacuum is described by the ground state wave functional. We must therefore solve for the ground state of the Hamiltonian in (\ref{H j}). The eigenstates of the Hamiltonian in (\ref{H j}) are well-known (see, for example, \cite{Ciftja1999}). To find the ground state, the first step is to solve the Ermakov-Milne-Pinney (EMP) equation, 
\begin{equation} \label{EMP eq}
\ddot{\rho}_{j}(\tilde{t}) + \frac{\dot{M}(\tilde{t})}{M(t)} \dot{\rho}_{j}(\tilde{t}) + \omega_{j}^2(\tilde{t}) \rho_{j}(\tilde{t})
= \frac{c^2_{j }}{M^2(\tilde{t}) \rho_{j}^3(\tilde{t})},
\end{equation}
for $\rho_{j}(\tilde{t})$ and $\dot{\rho}_{j}(\tilde{t})$, where a dot denotes a $\tilde{t}$-derivative and the $c_j$ are arbitrary constants (in practice, we set the $c_j$ equal to 1, but will retain them for completeness). Although the ground state requires only that $\rho_j$ satisfy the EMP equation and does not require boundary conditions, we are in need of boundary conditions for solving the EMP equation numerically. We use a static solution as an initial condition, defined by $\rho_j$ being time independent at $\tilde{t} = 0$,
\begin{equation} \label{EMP IC}
\rho_{j} = \left( \frac{c_{j}}{M \omega_j} \right)^{1/2},
\qquad
\dot{\rho}_{j} = 0,
\end{equation} 
where the first equation follows directly from (\ref{EMP eq}) for $\dot{\rho}_j = \ddot{\rho}_j = 0$ and where $M$ and $\omega_j$ are evaluated at $\tilde{t} = 0$. Once $\rho_j(\tilde{t})$ and $\dot{\rho}_j(\tilde{t})$ are found, the ground state for the $j$th oscillator is given by
\begin{equation}
\psi_{j, 0}(\tilde{t},y_j) = \left[\frac{c_{j}}{\pi \rho_{j}^2(\tilde{t})} \right]^{1/4}
\exp\left\{ - \frac{1}{2} \left[ 
\frac{c_{j}}{\rho_{j}^2(\tilde{t})} -  i \frac{M(\tilde{t})\dot{\rho}_{j}(\tilde{t})}{\rho_{j}(\tilde{t})}\right] y_j^2
\right\}
e^{-iE_{j,0} \tau_{j}(\tilde{t})},
\end{equation}
where
\begin{equation}
\tau_{j}(\tilde{t}) = \int \frac{d\tilde{t}}{M(\tilde{t}) \rho_{j}^2(\tilde{t})},
\qquad
E_{j,0} = \frac{c_j}{2}.
\end{equation}
Using (\ref{psi psi j}), the ground state is then
\begin{equation}  \label{psi 1}
\psi_0(\tilde{t}, \mathbf{y}) = \frac{\prod_{i=1}^N (c_i/\rho_i^2)^{1/4}}{\pi^{N/4}}
\exp\left( - \frac{1}{2}  \sum_{j,k = 1}^N 
y_j\Sigma_{jk} y_k
- i \sum_{j=1}^N E_{j,0} \tau_j
\right),
\end{equation}
where
\begin{equation} \label{Sigma def}
\Sigma_{jk}(\tilde{t}) = \left(\frac{c_j}{\rho_{j}^2} -  i \frac{M\dot{\rho}_{j}}{\rho_{j}} \right) \delta_{jk}
\end{equation}
is complex, time dependent, and diagonal. It is straightforward to decompose $\Sigma_{jk}$ into real and imaginary parts,
\begin{equation}
\Sigma_{jk} = \Sigma_{R,jk} + i \Sigma_{I,jk},
\qquad
\Sigma_{R,jk} = \frac{c_j}{\rho_{j}^2} \delta_{jk},
\qquad
\Sigma_{I,jk} = -\frac{M\dot{\rho}_{j}}{\rho_{j}} \delta_{jk}.
\end{equation}
It is not difficult to see that $\prod_{i=1}^N (c_i/\rho_i^2)^{1/4}/\pi^{N/4} = \left[ \det(\Sigma_R/\pi) \right]^{1/4}$.

The next step is to transform back to the original oscillators, $z_j$. We must do this in order to trace out the $z_j$ and form the reduced density matrix. The orthogonal matrix $U$ transforms the $z_j$ to the $y_j$, as can be seen from (\ref{y U z}). Transforming back to the $z_j$, the ground state becomes
\begin{equation} \label{psi 2}
\psi_0(\tilde{t},\mathbf{z}) = 
\left[ \det(\Sigma_R/\pi) \right]^{1/4}
\exp\left( - \frac{1}{2}  \mathbf{z}^T \Omega \mathbf{z} 
- i \sum_{k=j}^N E_{j,0} \tau_j
\right),
\end{equation}
where
\begin{equation} \label{Omega def}
\Omega(\tilde{t}) = U^T \Sigma U
\end{equation}
is complex, time dependent, and symmetric. The density matrix is given as usual by $\rho(\tilde{t};\mathbf{z}, \mathbf{z}') = \psi(\tilde{t},\mathbf{z})\psi^*(\tilde{t},\mathbf{z}')$. The ground state in (\ref{psi 2}) is then described by density matrix
\begin{equation} \label{density matrix}
\rho(\tilde{t}; \mathbf{z}, \mathbf{z}') 
= \left[ \det(\Sigma_R/\pi) \right]^{1/2}
\exp\left[ 
- \frac{1}{2} \left(  \mathbf{z}^T \Omega \mathbf{z} 
+ \frac{1}{2} \mathbf{z}^{\prime T} \Omega^* \mathbf{z}'
\right) \right].
\end{equation}
We find that the quantities $\tau_j$ and $E_{j,0}$ cancel out. In general, the entanglement entropy will not make use of them.

To construct the reduced density matrix, we trace out the first $n$ oscillators. More specifically, oscillators $1,2,\ldots n$ will be traced out and the reduced density matrix will describe oscillators $n+1, n+2, \ldots, N$. We have for the reduced density matrix,
\begin{equation} \label{red density matrix form}
\begin{split}
&\rho_\text{red}(\tilde{t}; z_{n+1}, \ldots z_N; z'_{n+1}, \ldots, z'_N) 
\\
\qquad&= 
\int  \left( \prod_{j=1}^n dz_j\right)
\rho(\tilde{t}; z_1, \ldots, z_n, z_{n+1}, \ldots, z_N; z_1, \ldots, z_n, z'_{n+1}, \ldots, z'_N).
\end{split}
\end{equation}
It is convenient to introduce the notation 
\begin{equation}
\mathbf{z} = 
\begin{pmatrix}
\mathbf{z}_n \\ \mathbf{z}_N
\end{pmatrix},
\qquad
\mathbf{z}_n = 
\begin{pmatrix}
z_1 \\ z_2 \\ \vdots \\ z_n
\end{pmatrix},
\qquad
\mathbf{z}_N = 
\begin{pmatrix}
z_{n+1} \\ z_{n+2} \\ \vdots \\ z_N
\end{pmatrix}
\end{equation}
and decompose $\Omega$ as
\begin{equation} \label{Omega block form}
\Omega = 
\begin{pmatrix}
\mathcal{A} & \mathcal{B} \\ \mathcal{B}^T & \mathcal{C}
\end{pmatrix},
\end{equation}
where $\mathcal{A}$ is $n\times n$ and symmetric, $\mathcal{C}$ is $(N-n)\times(N-n)$ and symmetric, and $\mathcal{B}$ is $n\times(N-n)$. The reduced density matrix can then be written
\begin{equation}
\begin{split}
\rho_\text{red}
&=  \left[ \det(\Sigma_R/\pi) \right]^{1/2}
\exp\left[ - \frac{1}{2} \left( \mathbf{z}_N^T \mathcal{C} \mathbf{z}_N
+ \mathbf{z}_N^{\prime T} \mathcal{C}^* \mathbf{z}'_N \right) \right]
\\
&\qquad
\times \int \left( \prod_{j=1}^n dz_j\right)
\exp\Bigl[ - \mathbf{z}_n^T \mathcal{A}_R \mathbf{z}_n
- (\mathcal{B}\mathbf{z}_N + \mathcal{B}^*\mathbf{z}'_N)^T \mathbf{z}_n \Bigr],
\end{split}
\end{equation} 
where $\mathcal{A}_R = \text{re}(\mathcal{A})$. The integral on the bottom line can be evaluated with standard formulas. Specifically,
\begin{equation}
\begin{split}
&\int \left( \prod_{j=1}^n dz_j\right)
\exp\Bigl[ - \mathbf{z}_n^T \mathcal{A}_R \mathbf{z}_n
- (\mathcal{B}\mathbf{z}_N + \mathcal{B}^*\mathbf{z}'_N)^T \mathbf{z}_n \Bigr]
\\
&\qquad
= \left[ \det (\pi \mathcal{A}_R^{-1}) \right]^{1/2}
\exp \left[
\frac{1}{4} (\mathcal{B}\mathbf{z}_N + \mathcal{B}^*\mathbf{z}'_N)^T \mathcal{A}_R^{-1} (\mathcal{B}\mathbf{z}_N + \mathcal{B}^*\mathbf{z}'_N)
\right].
\end{split}
\end{equation}
Before applying this formula, we define
\begin{equation} \label{gamma beta def}
\bar{\gamma} = \gamma + i \gamma_I \equiv  \mathcal{C} - \frac{1}{2}  \mathcal{B}^T \mathcal{A}_R^{-1} \mathcal{B},
\qquad
\beta \equiv \frac{1}{2} \mathcal{B}^\dag \mathcal{A}_R^{-1} \mathcal{B}.
\end{equation}
$\gamma$ and $\gamma_I$ are the real and imaginary parts of $\bar{\gamma}$, both of which are $(N-n)\times(N-n)$, real, and symmetric. $\beta$ is $(N-n)\times(N-n)$, complex, and hermitian. We have now for the reduced density matrix
\begin{equation} \label{rho red gamma delta}
\begin{split}
\rho_\text{red}
&= \left[ \det (\pi \mathcal{A}_R^{-1}) \right]^{1/2}
 \left[ \det(\Sigma_R/\pi) \right]^{1/2}
\exp\left( -\frac{1}{2} \mathbf{z}_N^T \gamma \mathbf{z}_N
-\frac{1}{2} \mathbf{z}_N^{\prime T} \gamma \mathbf{z}_N'
+ \mathbf{z}_N^{\prime T} \beta \mathbf{z}_N
\right)
\\
&\qquad \times
\exp \left( - \frac{i}{2} \mathbf{z}_N^T \gamma_I \mathbf{z}_N
+ \frac{i}{2} \mathbf{z}_N^{\prime T} \gamma_I \mathbf{z}_N' \right).
\end{split}
\end{equation}


\subsection{Eigenvalues}

In order to compute the entanglement entropy, we must determine the eigenvalues of the reduced density matrix. The eigenvalue equation for the reduced density matrix is 
\begin{equation} \label{rho red evalue eq}
\int \left( \prod_{i =n+1}^{N} dz_i' \right) 
\rho_\text{red}(\tilde{t}; \mathbf{z}_N, \mathbf{z}'_N) f(\tilde{t},\mathbf{z}'_N)
=
\lambda(\tilde{t}) f (\tilde{t},\mathbf{z}_N),
\end{equation}
where $\lambda$ are the eigenvalues and $f$ are the eigenfunctions. It is straightforward to show that the reduced density matrix in (\ref{rho red gamma delta}) and the reduced density matrix with $\gamma_I = 0$ have the same eigenvalues \cite{Boutivas:2023ksg, Katsinis:2023hqn} (but different eigenfunctions, which we will not be in need of). We therefore drop the $\gamma_I$ terms in (\ref{rho red gamma delta}) and compute the eigenvalues of 
\begin{equation} \label{rho red gamma delta 2}
\rho_\text{red}
= \left[ \det (\pi \mathcal{A}_R^{-1}) \right]^{1/2}
 \left[ \det(\Sigma_R/\pi) \right]^{1/2}
\exp\left( -\frac{1}{2} \mathbf{z}_N^T \gamma \mathbf{z}_N
-\frac{1}{2} \mathbf{z}_N^{\prime T} \gamma \mathbf{z}_N'
+ \mathbf{z}_N^{\prime T} \beta \mathbf{z}_N
\right).
\end{equation}

$\gamma$ is an $(N-n)\times(N-n)$ matrix that is real, symmetric, and positive definite. It can be diagonalized with an orthogonal matrix, $V$, where our convention is
\begin{equation} \label{gamma gamma_D}
\gamma = V^T \gamma_D V,
\qquad
\gamma_D = \gamma_i \delta_{ij},
\end{equation}
where $\gamma_D$ is the diagonalized matrix and $\gamma_i$ are the eigenvalues of $\gamma$. We define also
\begin{equation} \label{gamma V z}
\bar{z}_{N,i} = \sqrt{\gamma_i} 
\left(V \mathbf{z}_N \right)_i
\end{equation}
and
\begin{equation}
\bar{\beta}_{ij} =  
\frac{1}{\sqrt{\gamma_i}} 
\left( V \beta V^T\right)_{ij} 
\frac{1}{\sqrt{\gamma_j}},
\end{equation}
where $\bar{\beta}$ is $(N-n)\times(N-n)$, complex, and hermitian. In terms of these quantities, the reduced density matrix in (\ref{rho red gamma delta 2}) becomes
\begin{equation} \label{red density matrix 3}
\begin{split}
\rho_\text{red}
&= \left[ \det (\pi \mathcal{A}_R^{-1}) \right]^{1/2}
 \left[ \det(\Sigma_R/\pi) \right]^{1/2}
\exp\left(
- \frac{1}{2}  \bar{\mathbf{z}}_N^T \bar{\mathbf{z}}_N
- \frac{1}{2}  \bar{\mathbf{z}}_N^{\prime T}\bar{\mathbf{z}}'_N
+ \bar{\mathbf{z}}^{\prime T}_N \bar{\beta} \bar{\mathbf{z}}_N
\right).
\end{split}
\end{equation}
This density matrix is no longer normalized (i.e., its trace no longer equals 1). It is important that we normalize it. One can show that
\begin{equation}
\left[ \det (\pi \mathcal{A}_R^{-1}) \right]^{1/2}
 \left[ \det(\Sigma_R/\pi) \right]^{1/2}
=
\left[ \det(\gamma_D) \right]^{1/2}
\left\{
\frac{\det \left[ I - \text{re} (\bar{\beta}) \right]}
{\pi^{N-n}}
\right\}^{1/2}
\end{equation}
and that the normalized density matrix is obtained by dropping  $[\det(\gamma_D)]^{1/2}$. The normalized density matrix is then
\begin{equation} \label{red density matrix 4}
\begin{split}
\bar{\rho}_\text{red}
&= 
\left\{
\frac{\det \left[ I - \text{re} (\bar{\beta}) \right]}
{\pi^{N-n}}
\right\}^{1/2}
\exp\left(
 - \frac{1}{2} 
 \bar{\mathbf{z}}_N^T \bar{\mathbf{z}}_N
- \frac{1}{2}   \bar{\mathbf{z}}_N^{\prime T}\bar{\mathbf{z}}'_N
+ \bar{\mathbf{z}}^{\prime T}_N \bar{\beta} \bar{\mathbf{z}}_N
\right).
\end{split}
\end{equation}
Since the eigenvalue equation in (\ref{rho red evalue eq}) is invariant under any (non-singular) matrix transformations of $\mathbf{z}_N$, as long as the resulting density matrix is normalized, we may compute the eigenvalues from the eigenvalue equation
\begin{equation} \label{rho red evalue eq 2}
\int \left( \prod_{i =n+1}^{N} d\bar{z}_i' \right) 
\bar{\rho}_\text{red}(\tilde{t}; \mathbf{\bar{z}}_N, \mathbf{\bar{z}}'_N) f(\tilde{t},\mathbf{\bar{z}}'_N)
=
\lambda(\tilde{t}) f (\tilde{t},\mathbf{\bar{z}}_N).
\end{equation}

The $\bar{\beta}$ term in (\ref{red density matrix 4}) cannot be diagonalized with an orthogonal matrix, since $\bar{\beta}$ is complex and hermitian, and consequently the reduced density matrix cannot be diagonalized. (In contrast, for a static metric, $\bar{\beta}$ is real, symmetric, and can be diagonalized with a orthogonal matrix). Nevertheless, the eigenvalues of the reduced density matrix can still be computed. The derivation of how this is done is complicated and we refer to reader to \cite{Katsinis:2023hqn} for details. As shown in \cite{Katsinis:2023hqn}, the eigenvalues can be computed from the eigenvalue equation
\begin{equation} \label{eval eq}
\begin{pmatrix}
2\bar{\beta}^{-1} & - \bar{\beta}^{-1} \bar{\beta}^T
\\
I & 0
\end{pmatrix}
\begin{pmatrix}
\chi \\ \psi
\end{pmatrix}
= \lambda 
\begin{pmatrix}
\chi \\ \psi
\end{pmatrix}.
\end{equation}
This equation is written in block form, where each block of the matrix is $(N-n)\times(N-n)$, so that there are $2(N-n)$ eigenvalues. Performing matrix multiplication of the blocks and combining the two equations, (\ref{eval eq}) can be written as
\begin{equation} \label{quad evalue eq}
\left(2 I - \lambda \bar{\beta} - \frac{1}{\lambda} \bar{\beta}^T \right) \chi = 0.
\end{equation}
The eigenvalues are the solution to the equation $\det(2I - \lambda \bar{\beta} - \lambda^{-1} \bar{\beta}^T) = 0$. Since determinants are invariant under the transpose, the determinant equation defining the eigenvalues is invariant under the transformation $\lambda \rightarrow 1/\lambda$. This means for every eigenvalue $\lambda$, we have also $1/\lambda$ as an eigenvalue. We are only interested in those eigenvalues that are less than 1 \cite{Katsinis:2023hqn}. We label these eigenvalues $\xi_i$, $i = n+1,\ldots,  N$.

The equation for computing the eigenvalues can be simplified. First, (\ref{quad evalue eq}) can be rewritten as 
\begin{equation} \label{quadratic eigenvalue eq 2}
\left( \lambda^2 \beta   - 2\lambda \gamma + \beta^T   \right)
\zeta = 0,
\end{equation}
where $\zeta_i = (V^T \chi)_i / \sqrt{\gamma_i}$. In particular, $\beta$ is unbarred in (\ref{quadratic eigenvalue eq 2}). The quadratic eigenvalue equation in (\ref{quadratic eigenvalue eq 2}) can then be written as the generalized eigenvalue equation \cite{Belfiglio:2025hzo}
\begin{equation} \label{gen evalue eq}
F
\begin{pmatrix}
w \\ \zeta
\end{pmatrix}
= \lambda G
\begin{pmatrix}
w \\ \zeta
\end{pmatrix},
\qquad
F = 
\begin{pmatrix}
-2\gamma & \beta^T \\
I & 0
\end{pmatrix},
\qquad
G =
\begin{pmatrix}
-\beta & 0
\\
0 & I
\end{pmatrix},
\end{equation}
where $w = \lambda \zeta$. We will be computing the eigenvalues numerically. We have found that $F$ and $G$ can, at times, be close to singular, causing trouble for some numerical methods. This can be circumvented by introducing an arbitrary complex number $\sigma$ and solving the standard eigenvalue equation 
\begin{equation}
(M^{-1} G) v = \theta v,
\qquad
M \equiv F - \sigma G,
\end{equation}
for the eigenvalues $\theta = 1/(\lambda - \sigma)$. Once the $\theta$ are computed, the eigenvalues we are after are given by
\begin{equation}
\lambda = \sigma + \frac{1}{\theta}.
\end{equation}


\subsection{Entanglement entropy}

For given values of the quantum numbers $\ell$ and $m$, the entanglement entropy is  given by \cite{Boutivas:2023ksg, Katsinis:2023hqn}
\begin{equation}
S_\ell = - \sum_{i = n+1}^N
\left[
\ln  ( 1 - \xi_i)
+  \frac{\xi_i}{1-\xi_i} \ln \xi_i
\right].
\end{equation}
We have only included a subscript $\ell$ on $S_\ell$ because the entropy is independent of $m$.

The ground state is a product of ground states for each value of $\ell$ and $m$, as can be seen from (\ref{Psi psi lm}). As a consequence, so too is the density matrix. Since entropy is additive, $S(\rho_a \rho_b) = S(\rho_a) + S(\rho_b)$, we have for the total entanglement entropy
\begin{equation}
S(\tilde{t}, n) = \sum_{\ell = 0}^\infty (2\ell + 1) S_\ell.
\end{equation}
$S$ depends on the time, $\tilde{t}$, and the number of oscillators traced out, $n$, and hence on the region of space traced out.


\subsection{Procedure for computing the entanglement entropy}
\label{sec:procedure}

In this final subsection, we present the step-by-step procedure for computing the entanglement entropy of a discretized scalar field propagating in a fixed spacetime background described by the metric in (\ref{EE metric}). We have written this section to be stand-alone, in that it can be read without having read through the previous subsections. 

We discretize the $x$-axis as 
\begin{equation}
x_j = \left[ j - \frac{1}{2}(N + 1)\right]\epsilon,
\qquad
j = 1,2,\ldots,N,
\end{equation}
where $N$ is the number of lattice sites, $N$ must be even, and $\epsilon$ is the ultraviolet cutoff. This lattice is symmetric about $x = 0$ and does not contain $x = 0$. Points on the lattice with $x<0$ correspond to $j \leq N/2$ and points on the lattice with $x>0$ correspond to $j \geq N/2+1$. We will compute the entanglement entropy for the first $n$ oscillators traced out, i.e., oscillators $1,2,\ldots,n$ are traced out, leaving behind oscillators $n+1,n+2,\ldots,N$.

Spatially-dependent fields are discretized as
\begin{equation}
b_j = b(x_j), \qquad
r_j = r(x_j).
\end{equation}
We construct the matrix $K_{jk}$, where the diagonal components are 
\begin{equation}  \label{K diag}
\begin{split}
K_{11} &= -\beta_1
\\
K_{22} &= - \beta_2 - \alpha_1
\\
K_{33} &= - \beta_3 - \alpha_2
\\
&\vdots
\\
K_{N-1,N-1} &= -\beta_{N-1} - \alpha_{N-2}
\\
K_{N,N} &= -\beta_{N} - \alpha_{N-1}
\end{split}
\end{equation}
and the only off-diagonal components are
\begin{equation} \label{K off-diag}
K_{j,j+1} = K_{j+1,j} = -\gamma_j,
\qquad
1 \leq j \leq N-1,
\end{equation}
where
\begin{equation} 
\begin{split}
\alpha_j &= - \frac{\tilde{r}^2_{j+1/2}}{\tilde{r}^2_{j+1}}
\frac{1}{b_{j+1/2} b_{j+1}} 
\\
\beta_j &=
-\frac{\tilde{r}^2_{j+1/2}}{\tilde{r}_j^2}
\frac{ 1}{b_{j+1/2}b_j} 
-  \frac{\ell(\ell + 1)}{ \tilde{r}_j^2} 
\\
\gamma_j &=
\frac{\tilde{r}^2_{j+1/2}}{\tilde{r}_{j+1} \tilde{r}_j}
\frac{1}{b_{j+1/2}\sqrt{b_{j+1} b_{j}}}
\end{split}
\end{equation}
In these equations, $\tilde{r}_j = r_j/\epsilon$, $\tilde{\mu} = \mu \epsilon$, and $\ell = 0,1,\ldots$ is the standard quantum number that comes about from using spherical harmonics.

Next, we numerically compute the eigenvalues and eigenvectors of $K$. We use the eigenvectors to construct the diagonalizing orthogonal matrix $U$, defined by
$K = U^T K_D U$, where $(K_D)_{jk} = \Gamma_j^2 \delta_{jk}$ is the diagonal matrix and $\Gamma_i^2$ are the eigenvalues. 

We next construct
\begin{equation} 
M(\tilde{t}) = \frac{B^3(\tilde{t})}{A(\tilde{t})},
\qquad
\dot{M}(\tilde{t}) =  M(\tilde{t}) \left[ \frac{3 \dot{B}(\tilde{t})}{B(\tilde{t})} - \frac{\dot{A}(\tilde{t})}{A(\tilde{t})} \right],
\qquad
\omega_i^2(\tilde{t}) = \frac{A^2(\tilde{t})}{B^2(\tilde{t})} \left[ \Gamma_i^2 + B^2(\tilde{t}) \tilde{\mu}^2 \right],
\end{equation}
where $\tilde{t} = t/\epsilon$ and a dot denotes a $\tilde{t}$-derivative. We then numerically solve the Ermakov-Milne-Pinney (EMP) equation,\footnote{We numerically solve the EMP equation using the Dormand-Prince RK5(4) algorithm \cite{Press:2007ipz}.}
\begin{equation} 
\ddot{\rho}_{j}(\tilde{t}) + \frac{\dot{M}(\tilde{t})}{M(t)} \dot{\rho}_{j}(\tilde{t}) + \omega_{j}^2(\tilde{t}) \rho_{j}(\tilde{t})
= \frac{1}{M^2(\tilde{t}) \rho_{j}^3(\tilde{t})},
\end{equation}
using the initial conditions 
\begin{equation} 
\rho_{j} = \frac{1}{\sqrt{M \omega_j}},
\qquad
\dot{\rho}_{j} = 0,
\end{equation} 
where $M$ and $\omega_j$ are evaluated at $\tilde{t} = 0$.

With the solutions $\rho_j(\tilde{t})$ and $\dot{\rho}_j(\tilde{t})$ from the EMP equation, we construct the matrix
\begin{equation} 
\Sigma_{jk}(\tilde{t}) = \left(\frac{c_j}{\rho_{j}^2} -  i \frac{M\dot{\rho}_{j}}{\rho_{j}} \right) \delta_{jk}.
\end{equation}
We then compute
\begin{equation} 
\Omega(\tilde{t}) = U^T \Sigma U,
\end{equation}
which we decompose as
\begin{equation} 
\Omega = 
\begin{pmatrix}
\mathcal{A} & \mathcal{B} \\ \mathcal{B}^T & \mathcal{C}
\end{pmatrix},
\end{equation}
where $\mathcal{A}$ is $n\times n$, $\mathcal{C}$ is $(N-n)\times(N-n)$, and $\mathcal{B}$ is $n \times (N-n)$. We use these sub-matrices to compute 
\begin{equation} 
\gamma(\tilde{t}) = \text{re} \left(\mathcal{C} - \frac{1}{2}  \mathcal{B}^T \mathcal{A}_R^{-1} \mathcal{B}\right),
\qquad
\beta(\tilde{t}) = \frac{1}{2} \mathcal{B}^\dag \mathcal{A}_R^{-1} \mathcal{B},
\qquad
\mathcal{A}_R(\tilde{t}) = \text{re}(\mathcal{A}).
\end{equation}

We then numerically solve the generalized eigenvalue equation
\begin{equation} 
F
\begin{pmatrix}
w \\ \zeta
\end{pmatrix}
= \lambda G
\begin{pmatrix}
w \\ \zeta
\end{pmatrix},
\qquad
F = 
\begin{pmatrix}
-2\gamma & \beta^T \\
I & 0
\end{pmatrix},
\qquad
G =
\begin{pmatrix}
-\beta & 0
\\
0 & I
\end{pmatrix},
\end{equation}
for the eigenvalues $\lambda$. Some properties of $F$ and $G$ indicate it that it may be easier to instead solve the standard eigenvalue equation 
\begin{equation}
(M^{-1} G) v = \theta v,
\qquad
M \equiv F - \sigma G
\end{equation}
for the eigenvalues $\theta$, where $\sigma$ is arbitrary complex number. We then have $\lambda = \sigma + 1/\theta$.

We retain only those eigenvalues that are less than 1, the number of which is expected to be $N-n$. We label them $\xi_i$, $i = n+1,\ldots,  N$. The entanglement entropy is given by
\begin{equation}
S(\tilde{t}, n) = \sum_{\ell = 0}^\infty (2\ell + 1) S_\ell,
\qquad
S_\ell = - \sum_{i = n+1}^N
\left[
\ln  ( 1 - \xi_i)
+  \frac{\xi_i}{1-\xi_i} \ln \xi_i
\right].
\end{equation}
The summation over $\ell$ shows that we must perform this calculation for each value of $\ell = 0,1,\ldots$. Since we will be computing the entanglement entropy numerically, we cannot compute the contributions from the infinite values of $\ell$. Each term in the sum is generally smaller than the proceeding term. By comparing consecutive contributions, we can truncate the sum over $\ell$ once we reach a desired accuracy. For the results shown in the next section, we truncate the sum once the difference between consecutive values relative to their mean is less than $10^{-6}$.

In constructing these equations, we made use of the finite difference formulas defined in (\ref{finite diff}). Strictly speaking, these formulas are not symmetric about $x = 0$. While this leads to a small asymmetry in the computed entanglement entropy about $x = 0$, we have confirmed that our code computes an entanglement entropy that is consistent with the spacetime being symmetric about $x = 0$. To save computation time, we therefore compute the entanglement entropy for one side of the wormhole only. Numerically, it is more efficient to compute the entropy for $x > 0$ (i.e., tracing out oscillators $1,2,\ldots,n$ for $n \geq N/2+1$). When presenting results, we use the mirror image of the computed entropy for $x < 0$.


\section{Entanglement entropy of evolving wormholes}
\label{sec:EE evolving wormholes}

We are now ready to compute the entanglement entropy of a discretized scalar field in an evolving wormhole geometry. An evolving wormhole is described by the metric in (\ref{evolving metric}). As mentioned in section \ref{sec:evolving wormholes}, we will set $\Phi = 0$ and $h(r)$ equal to a constant, $h(r) = r_\text{th}$. A specific wormhole model is chosen by specifying the metric functions $A(\tilde{t})$ and $B(\tilde{t})$, where $A$ is subject to the evolving wormhole constraint in (\ref{A choices}), and $\tilde{r}_\text{th}$. In computing the entanglement entropy, we must also specify the scalar field mass $\tilde{\mu}$ and the number of lattice sites, $N$. Scaled quantities we will be making use of are defined by $\tilde{x} = x/\epsilon$, $\tilde{r} = r/\epsilon$, $\tilde{t} = t/\epsilon$, and $\tilde{\mu} = \mu \epsilon$, where $\epsilon$ is the UV cutoff defined in (\ref{grid def}). Length scales in the system, of course, should always be much larger than the UV cutoff. Since the minimum value of the areal radius is given by $r_\text{th}$, we require that $r_\text{th} \gg \epsilon$ or $\tilde{r}_\text{th} \gg 1$. From (\ref{grid def}),
\begin{equation} \label{eps x_N N}
N = 2\tilde{x}_N + 1,
\end{equation}
which informs us that a choice of $N$ is equivalent to a choice of the scaled comoving system size, $2\tilde{x}_N$. We always choose $N$ large enough that the presented results are robust and independent of the infrared cutoff. The presented results are therefore independent of $N$.

\begin{figure} 
\centering
\includegraphics[width=3.5in]{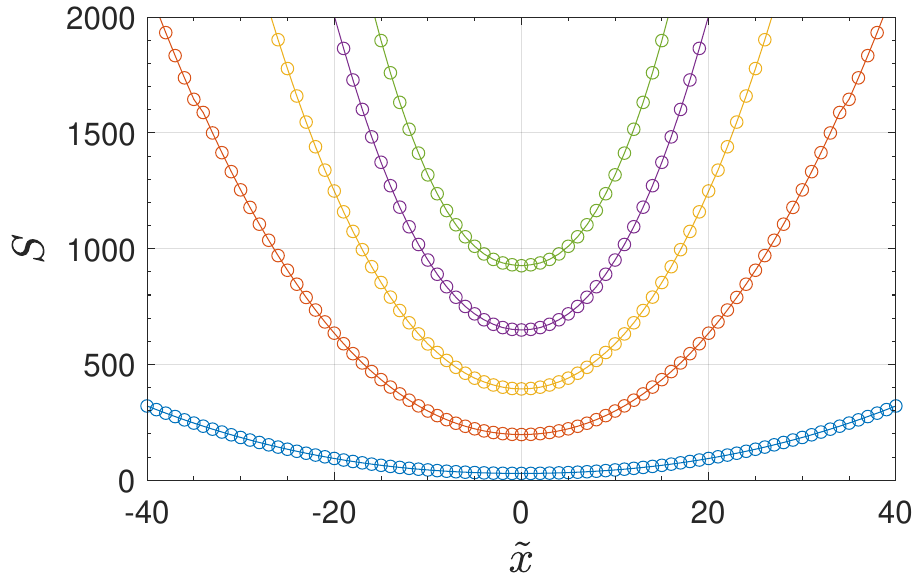} 
\caption{Entanglement entropy, $S$, as a function of the scaled comoving radial coordinate, $\tilde{x}$, for the expanding wormhole model defined in (\ref{e(+omega t)}) and parameter values $\tilde{r}_\text{th} = 10$, $\tilde{\omega} = 1$, and $\tilde{\mu} = 0$. From bottom to top, the curves are for $\tilde{t} = 0$, 1, 2, 3, and 4. The circles are the computed values which have been connected by straight lines.}
\label{fig:1}
\end{figure}

We will focus on two (related) models. The first model we consider is defined by
\begin{equation} \label{e(+omega t)}
A(\tilde{t}) = B(\tilde{t}) = e^{\tilde{w}\tilde{t}},
\end{equation}
where $\tilde{\omega} = \epsilon \omega$ is a positive constant. This model, which describes an expanding wormhole, was studied in \cite{Kar:1994tz, Kar:1995ss}, where it was shown to satisfy the weak energy condition. For simplicity, we set $\tilde{r}_\text{th} = 10$, $\tilde{\omega} = 1$, and $\tilde{\mu} = 0$. We begin by presenting $S$ as a function of the scaled comoving radial coordinate $\tilde{x}$. We worked out how to compute $S(\tilde{t},n)$ when tracing out lattice sites $1,2,\ldots,n$ in section \ref{sec:EE} and for each $n$ we have the corresponding value $\tilde{x}_n$. For the purposes of plotting, when presenting the entanglement entropy $S(\tilde{t}, n)$ we will take the corresponding radial coordinate and areal radius to be $\tilde{x}_{n+1/2}$ and $\tilde{r}_{n+1/2}$, which approximates the boundary between lattice sites $j=n$ and $j= n +1$. In figure \ref{fig:1}, we show results for, from bottom to top, $\tilde{t} = 0$, 1, 2, 3, and 4. The circles are the computed values, which we have connected by straight lines. We find that the entanglement entropy has a minimum at the center of the wormhole. As time increases and the wormhole expands, the entanglement entropy is seen to increase.  

While the plot in figure \ref{fig:1} shows the entropy increasing with time, we must be careful before we interpret this growth as being caused by genuine dynamically generated entropy. First, the increase could be an artifact of the expanding conformal factor in (\ref{e(+omega t)}), which in effect is a relabeling of the coordinate grid. However, figure \ref{fig:1} is plotted in terms of the comoving radial coordinate $\tilde{x}$, which is independent of the conformal factor, allowing us to rule out coordinate expansion as the cause. Second, we must consider the evolution of the physical UV cutoff length scale, $\epsilon_\text{phys} = B(\tilde{t}) \epsilon$. As the physical cutoff expands, the cutoff energy scale decreases and we effectively probe fewer and fewer degrees of freedom, which we might expect to lead to a decrease in measured entropy. Despite this cutoff induced suppression, we find the entropy is increasing. We can therefore conclude that the increase of entropy in our result is dynamically generated.

We now study whether our results satisfy the area law \cite{Bombelli:1986rw, Srednicki:1993im, Solodukhin:2011gn},
\begin{equation} \label{area law}
S = 
c_1 \frac{A_\text{phys}}{\epsilon_\text{phys}^2}
+ c_2 \ln \left(\frac{A_\text{phys}}{\epsilon_\text{phys}^2} \right) + \cdots,
\end{equation}
where the omitted terms are finite in the limit $\epsilon_\text{phys} \rightarrow 0$. The first term in (\ref{area law}) is the leading order contribution  and the second term is the logarithmic correction. Both terms are written in terms of physical quantities. However,
\begin{equation} \label{A/eps phys}
\frac{A_\text{phys}}{\epsilon_\text{phys}^2}
= \frac{4\pi [B(\tilde{t}) r ]^2}
{[B(\tilde{t}) \epsilon]^2}
= 4\pi \left( \frac{r}{\epsilon} \right)^2
= 4\pi \tilde{r}^2
\end{equation}
and the area law can be written
\begin{equation}  \label{area law 2}
\begin{split}
S 
&= c_1 4\pi \tilde{r}^2
+ c_2 \ln (4\pi \tilde{r}^2) + \cdots.
\end{split}
\end{equation}
Since the conformal factor cancels out, the ratio $A_\text{phys}/\epsilon_\text{phys}^2$ is in fact time independent. Conveniently, the area law can be written exclusively in terms of comoving quantities.

\begin{figure} 
\centering
\includegraphics[width=3 in]{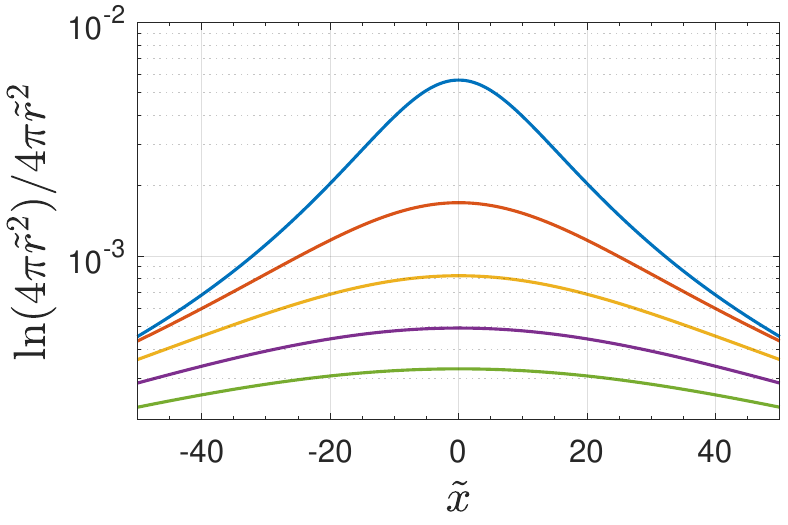} 
\caption{The logarithmic correction in the area law, for the area law as written in (\ref{area law 3}), is plotted as a function of scaled comoving radial coordinate $\tilde{x}$. This plot applies specifically to the wormhole model under study, where the relationship between $\tilde{x}$ and $\tilde{r}$ is given by (\ref{x(r) tilde}). From top to bottom, the curves are for $\tilde{r}_\text{th} = 10$, 20, 30, 40, and 50.}
\label{fig:2}
\end{figure}

It is helpful to reorganize the area law in (\ref{area law 2}) to
\begin{equation} \label{area law 3}
\frac{S}{4\pi \tilde{r}^2}
= c_1 + c_2 \frac{\ln (4\pi \tilde{r}^2)}{4\pi \tilde{r}^2}
+\cdots.
\end{equation}
If we plot $S/4\pi \tilde{r}^2$ and the result is constant (for a given time), we can conclude that the entanglement entropy is dominated by the leading order term. If, on the other hand, the result is not constant, then higher order terms are significant and must be taken into account to describe the entanglement entropy's scaling behavior. Before presenting results, we can get a sense for the logarithmic correction. $\ln(4\pi \tilde{r}^2)/4\pi\tilde{r}$ is time independent and, for the models we are considering, $\tilde{x}(\tilde{r})$ is known analytically, 
\begin{equation} \label{x(r) tilde}
\tilde{x} = \pm \tilde{r} \left( \sqrt{1 - \frac{\tilde{r}_\text{th}}{\tilde{r}}} 
+ \frac{\tilde{r}_\text{th}}{\tilde{r}} \tanh^{-1} \sqrt{1 - \frac{\tilde{r}_\text{th}}{\tilde{r}}}
\right),
\end{equation}
which follows directly from (\ref{x(r)}). In figure \ref{fig:2}, we show $\ln(4\pi \tilde{r}^2)/4\pi\tilde{r}^2$ as a function of $\tilde{x}$ for, from top to bottom, $\tilde{r}_\text{th} = 10$, 20, 30, 40, and 50. We find that the logarithmic correction has the best chance of being relevant near the center of the wormhole and for smaller values of $\tilde{r}_\text{th}$ and that we should expect it to be negligible far from the center of the wormhole at large $|\tilde{x}|$.

\begin{figure} 
\centering
\includegraphics[width=6 in]{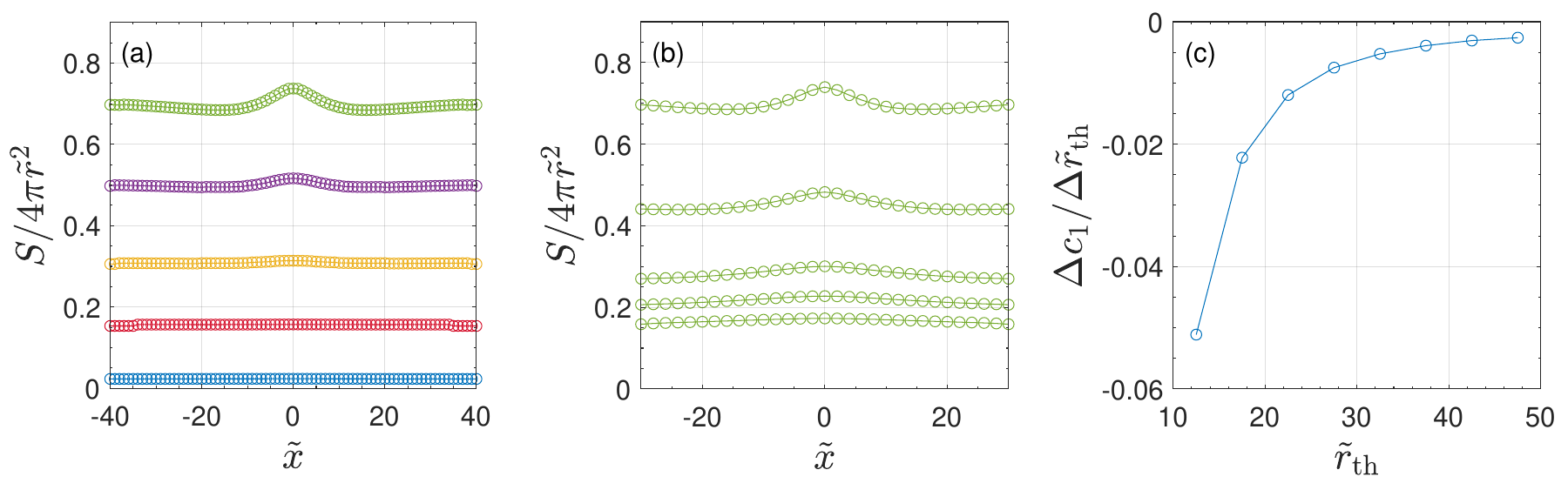} 
\caption{(a) The results shown in figure \ref{fig:1} are plotted here as $S/4\pi\tilde{r}^2$. (b) The top curve is the same as the top curve in (a). From top to bottom, the curves are for $\tilde{r}_\text{th} = 10$, 15, 25, 35, and 50, while keeping $t$ and $\omega$ (without tilde) unchanged. We may interpret these results as studying the same wormhole model while decreasing the value of the UV cutoff. (c) For the curves shown in (b) (as well as additional curves not shown), we display the rate of change of the area law coefficient $c_1$, as defined in (\ref{area law 3}), in terms of the increasing values of $\tilde{r}_\text{th}$.}
\label{fig:3}
\end{figure}

In figure \ref{fig:3}(a), we show $S/4\pi\tilde{r}^2$ as a function of $\tilde{x}$ for the same results shown in figure \ref{fig:1}. We see again that entropy increases with increasing time, since $S/4\pi\tilde{r}^2$ is increasing and $\tilde{r}$ is time independent. At early times, the curves are flat and the leading order term dominates the area law everywhere. Near the center of the wormhole, as time increases, the curves become markedly non-flat, indicating that higher order terms cannot be ignored in the area law in this region. 

These results have an interesting dependence on the UV cutoff $\epsilon$. By decreasing $\epsilon$, we can approach the continuum limit. In doing this, we are careful to analyze the same wormhole model by fixing $r_\text{th}$ and $\omega$ (without tildes). We are also careful to analyze the system at the same time $t$ (without tilde). Since $t/r_\text{th} = \tilde{t}/\tilde{r}_\text{th}$ and $\omega r_\text{th} = \tilde{\omega} \tilde{r}_\text{th}$, we have that $\tilde{t}/\tilde{r}_\text{th}$ and $\tilde{\omega} \tilde{r}_\text{th}$ are fixed and
\begin{equation} \label{tilde t omega}
\tilde{t} = \left(\frac{\tilde{t}}{\tilde{r}_\text{th}}\right)_\text{init} \tilde{r}_\text{th},
\qquad
\tilde{\omega} = ( \tilde{\omega} \tilde{r}_\text{th} )_\text{init}
\frac{1}{\tilde{r}_\text{th}}.
\end{equation}
In practice, we choose initial values for $\tilde{t}$, $\tilde{\omega}$, and $\tilde{r}_\text{th}$, which gives $(\tilde{t}/\tilde{r}_\text{th})_\text{init}$ and $(\tilde{\omega} \tilde{r}_\text{th})_\text{init}$. We then increase $\tilde{r}_\text{th}$ and use (\ref{tilde t omega}) to determine the values for $\tilde{t}$ and $\tilde{\omega}$ that we should use when computing the entropy. We interpret the increasing value of $\tilde{r}_\text{th}$ as a consequence of decreasing the UV cutoff $\epsilon$. We show results in figure \ref{fig:3}(b). The initial values are $\tilde{t} = 4$, $\tilde{r}_\text{th} = 10$, and $\tilde{\omega} = 1$. The top curve has $\tilde{r}_\text{th} = 10$ (making the top curve the same as the top curve in figure \ref{fig:3}(a)). Moving down, $\tilde{r}_\text{th} = 15$, 25, 35, and 50. As we decrease $\epsilon$, we find the curves lowering. This lowering has ramifications for computing the area law coefficient $c_1$ defined in (\ref{area law 3}), as we now explain.

\begin{figure} 
\centering
\includegraphics[width=6in]{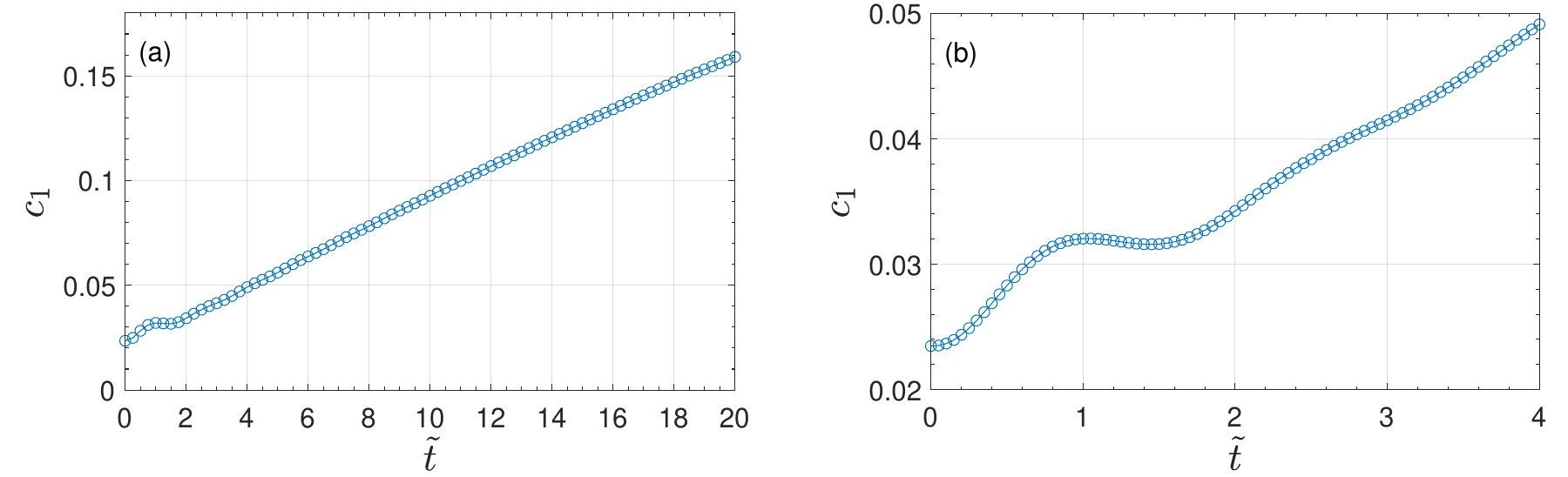} 
\caption{(a) The area law coefficient $c_1$ for the wormhole model with  $\tilde{r}_\text{th} = 50$ and $\tilde{\omega} = 0.2$. These same parameters and $\tilde{t} = 20$ are used to compute the bottom curve in figure \ref{fig:3}(b). (b) This plot zooms into the early time region in (a).}
\label{fig:4}
\end{figure}

From figure \ref{fig:2}, we know that the logarithmic correction to the area law is negligible at large $|\tilde{x}|$. We can therefore straightforwardly compute the area law coefficient $c_1$ by computing $S/4\pi \tilde{r}^2$ at large $|\tilde{x}|$. We display $\Delta c_1 / \Delta \tilde{r}_\text{th}$ in figure \ref{fig:3}(c), which measures the rate of change of $c_1$ as we increase $\tilde{r}_\text{th}$, for the same wormhole model used in figure \ref{fig:3}(b). Although $c_1$ continues to change over the numerically accessible range, the rate of change becomes progressively smaller in magnitude. Figure \ref{fig:3}(c) indicates that $c_1$ approaches a cutoff-independent value in the continuum limit. Figure \ref{fig:3}(c) also indicates that to obtain a robust value for $c_1$, we may have to approach the continuum limit. The bottom curve in figure \ref{fig:3}(b) is computed using $\tilde{r}_\text{th} = 50$, $\tilde{\omega} = 0.2$, and $\tilde{t} = 20$. Using the same values for $\tilde{r}_\text{th}$ and $\tilde{\omega}$, we show $c_1$ for $0\leq \tilde{t} \leq 20$ in figure \ref{fig:4}(a). Figure \ref{fig:4}(b) zooms into the region $0\leq \tilde{t} \leq 4$. We find that after some initial nonlinear transient behavior, the system settles into a linear relationship between $c_1$ and $\tilde{t}$.

Understanding the higher order terms in the area law is complicated, especially near the center of the wormhole. Figures \ref{fig:3}(a) and \ref{fig:3}(b) show that the entanglement entropy profiles exhibit a double-dip structure---a local maximum at the center flanked by two local minima---which cannot be accounted for by the logarithmic correction, which has no such minima, as can be seen from figure \ref{fig:2}. We therefore do not attempt to extract a continuum value of $c_2$ from an analysis of this region. On the other hand, there may be a window of $\tilde{x}$ values which is sufficiently far from the center that terms higher order than the logarithmic correction are negligible, but not so far from the center that the logarithmic correction is negligible. In this window it might be possible to compute the coefficient $c_2$ in the area law. Establishing this appears to require greater numerical accuracy than used in this work. More generally, subleading entanglement entropy terms are known to be
sensitive to corrections to asymptotic scaling \cite{Cardy:2010zs}
and to geometric effects \cite{Solodukhin:2011gn}. We therefore leave both the origin of the near-throat structure and the determination of the asymptotic subleading correction for future work.

An important question is whether or not the entanglement entropy obeys the area law. At early times, the answer is yes, as can be seen by the flat curves on the bottom of figure \ref{fig:3}(a). At later times, the answer may depend on the UV cutoff. More specifically, the answer could again be yes in the continuum limit, as indicated by the flatness of the lower curves in figure \ref{fig:3}(b), though establishing this would require computing to much larger values of $\tilde{x}$. In the discrete limit, however, the leading and logarithmic contributions are insufficient for describing the entropy near the center of the wormhole, but it may be possible for them to describe the entropy asymptotically far from the center. 

\begin{figure} 
\centering
\includegraphics[width=6 in]{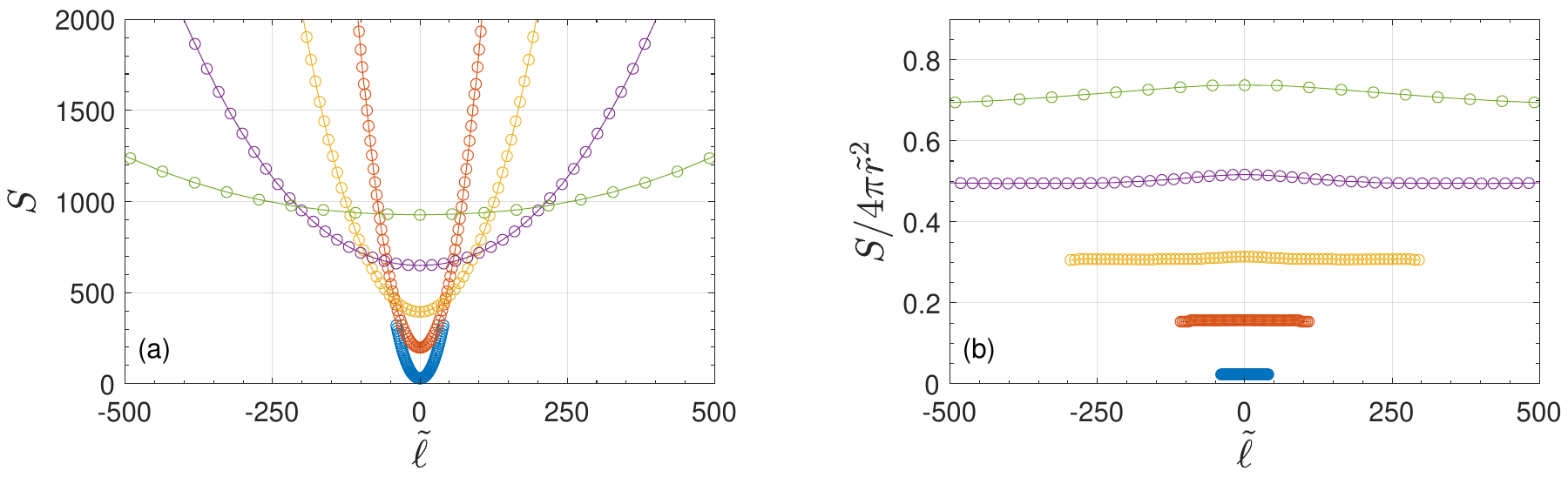} 
\caption{(a) The same results shown in figure \ref{fig:1} and (b) the same results shown in figure \ref{fig:3}(a), but now with both plots in terms of the scaled physical distance from the center of the wormhole, $\tilde{\ell}$. All curves in both plots extend indefinitely far out, but for an expanding spacetime it is numerically time consuming to compute the early times curves to the edge of the plot.}
\label{fig:5}
\end{figure}

We now consider the observations of an observer. Observers can measure physical quantities, such as the proper distance defined in (\ref{proper dist}), but cannot measure comoving quantities. In figure \ref{fig:5}(a), we display the same results shown in figure \ref{fig:1}, but now in terms of the scaled physical distance from the center of the wormhole, $\tilde{\ell}$. (We note that the blue curve continues indefinitely, but that it is numerically time consuming to compute this curve to the edge of the plot.) Since $\tilde{\ell} = B(\tilde{t})\tilde{x}$ and $B(\tilde{t})$ increases exponentially with time, the curves get stretched horizontally as time increases compared to figure \ref{fig:1}. This leads to the curves crossing one another. More quantitatively, we can consider the area law in (\ref{area law 2}). Sufficiently far from the throat, where $\tilde{r}\approx|\tilde{x}|$ and higher order contributions are negligible, the area law can be approximated by
\begin{equation} \label{S large |x|}
S(\tilde{t},\tilde{x}) \approx 4\pi c_1(\tilde{t})\tilde{x}^2.
\end{equation}
Using $\tilde{\ell}=B(\tilde{t})\tilde{x}$, this becomes
\begin{equation}
	S(\tilde{t},\tilde{\ell})
	\approx
	4\pi
	\frac{c_1(\tilde{t})}{B^2(\tilde{t})}
	\tilde{\ell}^{\,2}.
\end{equation}
Far from the center of the wormhole, the time dependence of the entanglement entropy is controlled by $c_1(\tilde{t})/B^2(\tilde{t})$. As mentioned, $B(\tilde{t})$ increases exponentially with time, while $c_1(\tilde{t})$ in general changes much more mildly, as can be see in figure \ref{fig:4}, leading to a decrease in entropy with time.

A consequence of the curves crossing in figure $\ref{fig:5}(a)$ is that an observer making such a measurement would mistakenly think that entropy is decreasing with time, at least away from the wormhole's center. The takeaway is that this is a poor choice for a measurement. A more sensible choice is shown in figure \ref{fig:5}(b). We showed in (\ref{A/eps phys}) that $4\pi \tilde{r}^2 = A_\text{phys}/\epsilon_\text{phys}^2$ and so this figure is properly parameterized in terms of physical quantities. (We note that all curves in figure \ref{fig:5}(b) extend indefinitely far out, but that it is numerically time consuming to compute the early time curves out to the edge of the plot.) An observer making this measurement would correctly deduce that entropy is increasing with time.

\begin{figure} 
\includegraphics[width=6in]{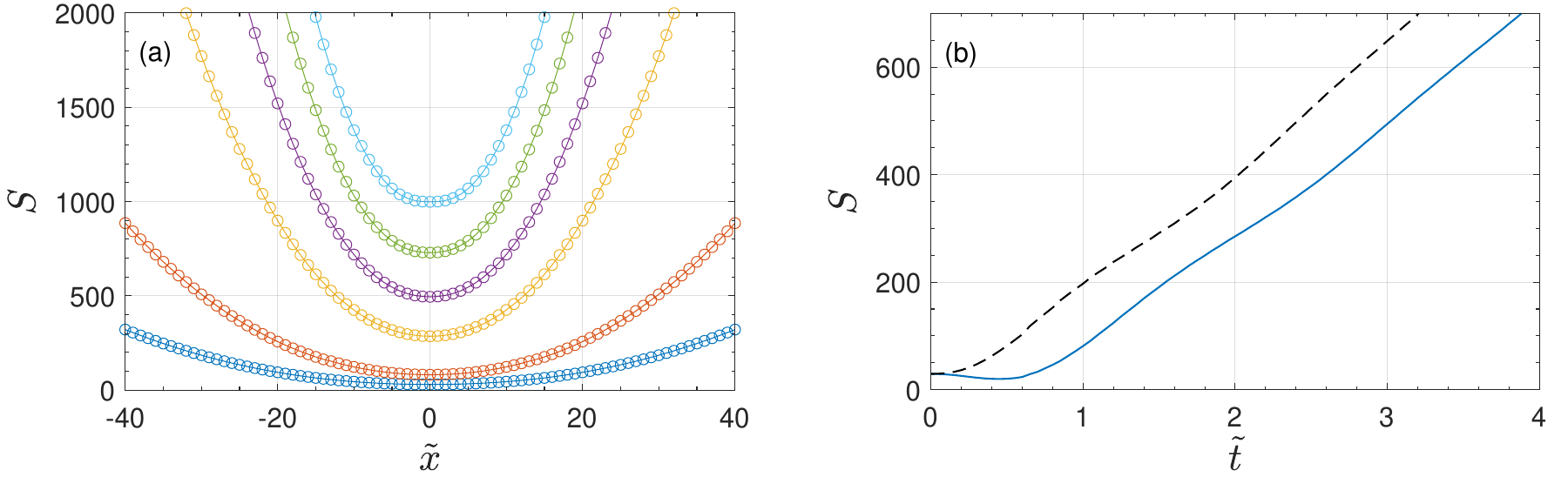} 
\caption{(a) Entanglement entropy, $S$, as a function of the scaled comoving radial coordinate, $\tilde{x}$, for the contracting wormhole model defined in (\ref{e(-omega t)}) and parameter values $\tilde{r}_\text{th} = 10$, $\tilde{\omega} = 1$, and $\tilde{\mu} = 0$. From bottom to top, the curves are for $\tilde{t} = 0$, 1, 2, 3, and 4. (b) The solid blue curve displays the entanglement entropy at $\tilde{x} = 0$ as a function of time for the same model shown in (a). We can see an initial dip before the entropy starts increasing. For comparison, we include the dashed black curve, which displays the entanglement entropy at $\tilde{x} = 0$ for the expanding model shown in figure \ref{fig:1}.}
\label{fig:6}
\end{figure}

We now consider a second (related) model, defined by
\begin{equation} \label{e(-omega t)}
A(\tilde{t}) = B(\tilde{t}) = e^{-\tilde{w}\tilde{t}},
\end{equation}
where $\tilde{\omega}$ is again a positive constant. Compared to the first model in (\ref{e(+omega t)}), this model has a negative sign in the exponent and describes a contracting wormhole. This model also satisfies the weak energy condition \cite{Kar:1994tz, Kar:1995ss}. As with the first model, we set $\tilde{r}_\text{th} = 10$, $\tilde{\omega} = 1$, and $\tilde{\mu} = 0$. We show the entanglement entropy as a function of the scaled comoving radial coordinate in figure \ref{fig:6}(a), just as we did for the first model in figure \ref{fig:1}. We can again conclude that the perceived increase in entropy cannot be attributed solely to a relabeling of the grid (in this case, a coordinate contraction), since we find the entropy increasing when plotted in terms $\tilde{x}$, which is independent of the conformal factor. However, with a contracting spacetime, the physical UV cutoff energy scale is increasing. As a consequence, more and more degrees of freedom are being probed and we might expect to find an increase in entropy, which is consistent with figure \ref{fig:6}(a). Unlike with the first model, we cannot conclude from figure \ref{fig:6}(a) that the increase in entropy is dynamically generated. We will show soon how this determination can be made.

Although figure \ref{fig:6}(a) shows the entropy to be increasing, it seems to do so more slowly at early times. In fact, the entropy undergoes a brief decrease before increasing. We show in figure \ref{fig:6}(b) as the solid blue curve the entanglement entropy at $\tilde{x} = 0$ as a function of time for the same model shown in figure \ref{fig:6}(a). At early times, we can see a slight dip before it begins increasing. For comparison, we show as the dashed black curve the entanglement entropy at $\tilde{x} = 0$ for the expanding model shown in figure \ref{fig:1}. We note that a similar dip was seen in \cite{Belfiglio:2025hzo} for the entanglement entropy in the contracting Oppenheimer-Snyder model, which describes a matter system collapsing to a black hole.

\begin{figure} 
\includegraphics[width=\textwidth]{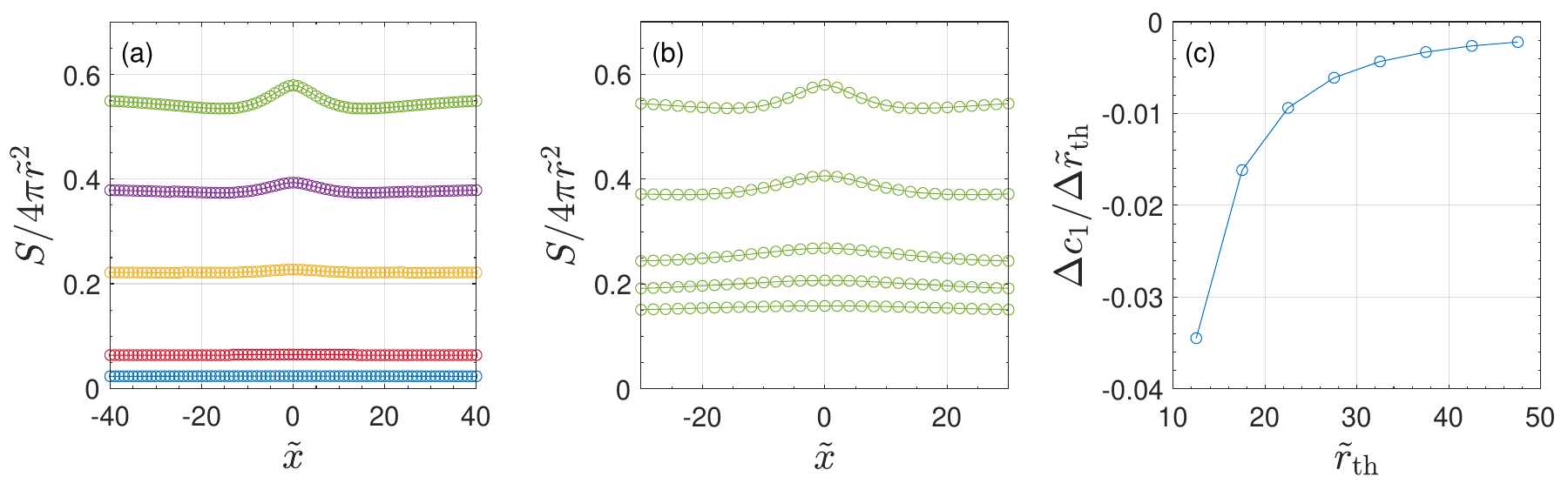} 
\caption{These plots are analogous to figure \ref{fig:3}, but for the contracting wormhole model. (a) The same results shown in figure \ref{fig:6}(a) are plotted here as $S/4\pi\tilde{r}^2$. (b) The top curve is the same as the top curve in (a). From top to bottom, the curves are for $\tilde{r}_\text{th} = 10$, 15, 25, 35, and 50, while keeping $t$ and $\omega$ (without tilde) unchanged. We may interpret these results as studying the same wormhole model while decreasing the value of the UV cutoff. (c) For the curves shown in (b) (as well as additional curves not shown), we display the rate of change of the area law coefficient $c_1$, as defined in (\ref{area law 3}), in terms of the increasing values of $\tilde{r}_\text{th}$.}
\label{fig:7}
\end{figure}

\begin{figure} 
\centering
\includegraphics[width=6in]{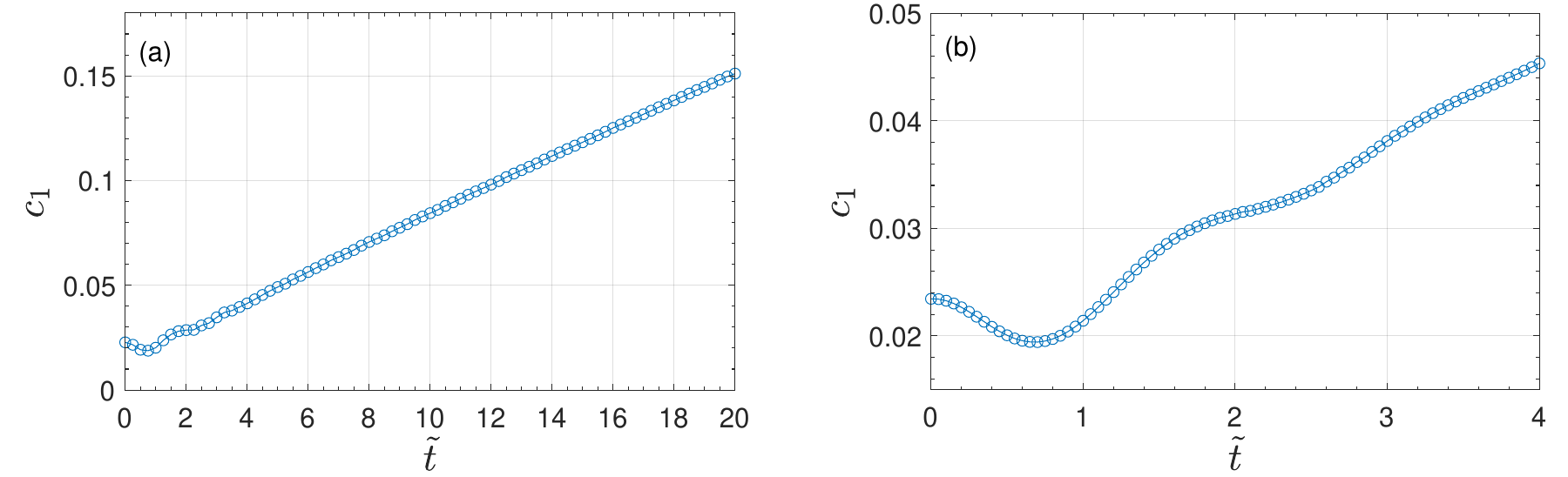} 
\caption{(a) The area law coefficient $c_1$ for the wormhole model with  $\tilde{r}_\text{th} = 50$ and $\tilde{\omega} = 0.2$. These same parameters and $\tilde{t} = 20$ are used to compute the bottom curve in figure \ref{fig:7}(b). (b) This plot zooms into the early time region in (a). Note the initial dip in $c_1$, which is consistent with the dip seen in the solid blue curve in figure \ref{fig:6}(b).}
\label{fig:8}
\end{figure}

We now consider the area law, where we find results very similar to the first model. In figure \ref{fig:7}(a), we show $S/4\pi \tilde{r}^2$ as a function of $\tilde{x}$. At early times, the curves are flat and the system is described well by the dominant contribution to the area law. At later times near the center, the curves are no longer flat and exhibit the double-dip structure, indicating that higher order terms in the area law are important. Using initially the same wormhole system as used for the top green curve in figure \ref{fig:7}(a), we decrease the UV cutoff $\epsilon$ (by increasing $\tilde{r}_\text{th}$). The results are shown in figure \ref{fig:7}(b), where we move down the curves as $\epsilon$ is decreased. Approaching the continuum limit, the entropy curves lower. We show in figure \ref{fig:7}(c) the rate of change of the area law coefficient $c_1$ for the results shown in figure \ref{fig:7}(b), which indicates that we approach a cutoff independent value of $c_1$ as we approach the continuum limit. Using the same parameters as used for the bottom curve in figure \ref{fig:7}(b), we show $c_1$ as a function of $\tilde{t}$ in figures \ref{fig:8}(a) and \ref{fig:8}(b). We again find nonlinear transient behavior before a linear relationship between $c_1$ and $\tilde{t}$ forms.

Just as in (\ref{S large |x|}), we assume that $S(\tilde{t},\tilde{x})/4\pi\tilde{r}^2 \approx c_1(\tilde{t})$ holds at large $|\tilde{x}|$, as follows from the area law. The use of $4\pi \tilde{r}^2$ is significant because it takes into account the decreasing length of the physical UV cutoff through (\ref{A/eps phys}), yet is conveniently written in terms of comoving quantities. Since we find that entanglement entropy increases in figure \ref{fig:7}(a), we can conclude, at least at large $|\tilde{x}|$, that the entropy growth is independent of the contracting UV cutoff and is in fact dynamically generated.

\begin{figure} 
\centering
\includegraphics[width=6 in]{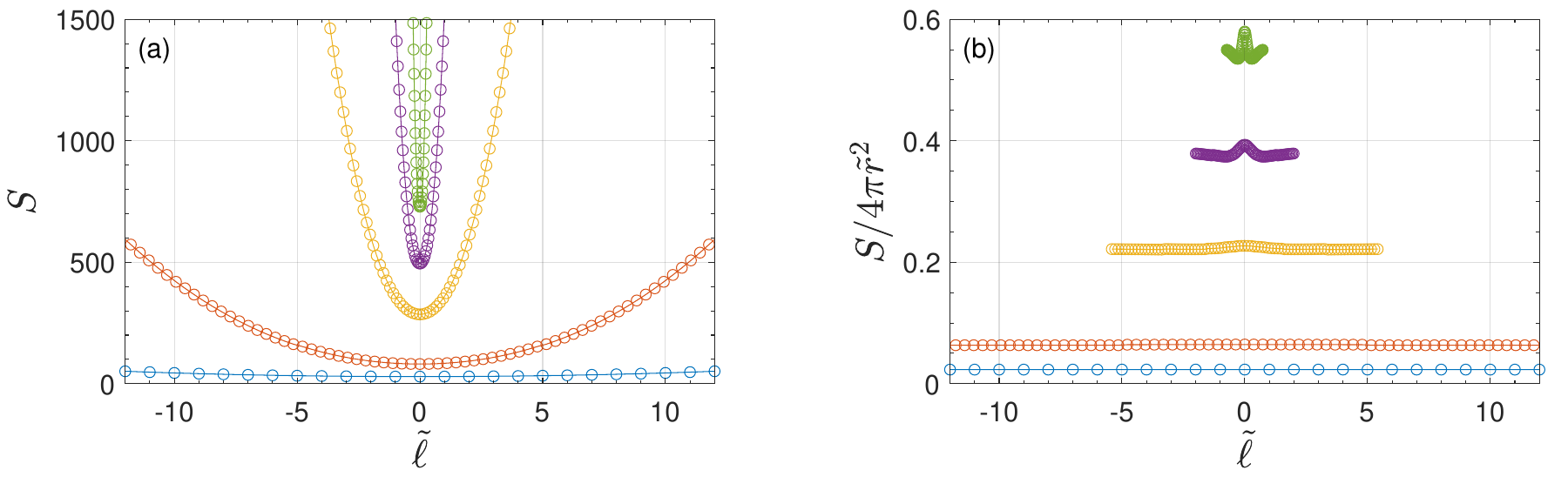} 
\caption{(a) The same results shown in figure \ref{fig:6}(a) and (b) the same results shown in figure \ref{fig:7}(a), but now with both plots in terms of the scaled physical distance from the center of the wormhole, $\tilde{\ell}$. For (b), all curves extend indefinitely far out, but for a contracting spacetime it is numerically time consuming to compute the late times curves to the edge of the  plot. }
\label{fig:9}
\end{figure}

In figures \ref{fig:9}(a) and \ref{fig:9}(b), we show $S$ and $S/4\pi\tilde{r}^2$ as functions of the physical distance $\tilde{\ell}$. Both of these plots show quantities that are measurable by an observer. An observer making such measurements would deduce that entropy is increasing with time, though we remind that it is only the measurement in figure \ref{fig:9}(b) that would lead to the correct conclusion for the expanding wormhole model. (We note that all curves in figure \ref{fig:9}(b) extend indefinitely, but that it is numerically time consuming to compute the late time curves out to the edge of the plot.)

\section{Conclusion}
\label{sec:conclusion}

We computed the entanglement entropy of a discretized scalar field in
the fixed spacetime background of an evolving wormhole. We focused on
two related models, one describing an expanding wormhole and the other
a contracting wormhole. Despite the opposing types of evolution in
these models, we found qualitatively similar behavior for the
entanglement entropy.

Our first main result is that the evolution of the wormhole dynamically
generates entanglement. For the expanding model, the increase in
entanglement entropy cannot be attributed to the expanding coordinate
grid, since the increase is present when the entropy is expressed in
terms of the comoving radial coordinate. Furthermore, the physical UV
cutoff increases during the expansion, which by itself would tend to
reduce the measured entropy. Nevertheless, the entropy increases with
time. For the contracting model, the decreasing physical UV cutoff
makes the interpretation of the unnormalized entropy less direct.
However, after accounting for the changing physical UV cutoff through
the area law normalization, we again find entropy growth that is
independent of the changing cutoff. We therefore find genuine
dynamical generation of entanglement in both expanding and contracting
wormhole geometries. The contracting model additionally exhibits a
brief initial decrease before the subsequent growth.

Our second main result concerns the area law. Asymptotically far from the wormhole throat, the entanglement entropy is well-described by the leading contribution to the area law. At finite UV cutoff, the corresponding
coefficient $c_1$ depends on the cutoff. By systematically decreasing
the UV cutoff, however, we find that the rate of change of $c_1$
becomes progressively smaller in magnitude. This occurs for both the
expanding and contracting models and provides numerical evidence that
$c_1$ approaches a cutoff-independent value in the continuum limit. 

The behavior near the wormhole throat is more complicated. At later
times in the evolution, the entropy profiles develop a characteristic double-dip structure, consisting of a local maximum at the throat flanked by two
local minima. This structure cannot be reproduced by the leading and
logarithmic contributions to the area law.  This does not exclude the possibility that the logarithmic correction remains applicable sufficiently far from the throat. Determining the asymptotic subleading correction would require greater numerical accuracy, and we leave both this question and the
origin of the near-throat structure for future work.

We have presented results for two closely related evolving wormhole
models. Initial studies of other evolving wormhole models, including models with nonzero mass for the scalar field, show
qualitatively similar behavior, including the initial dip for
contracting models and the appearance of the double-dip structure near
the throat at later times. It remains to be determined how general
these features are and whether similar behavior occurs for more general
time dependent wormhole geometries. More broadly, evolving wormholes
provide a useful setting in which the dynamical generation of quantum
entanglement, its leading area law behavior, and additional
geometry sensitive structure can be studied within the same
calculation.



\providecommand{\href}[2]{#2}\begingroup\raggedright\endgroup


\end{document}